\documentclass{kodra_ametsoc}
\journal{jcli}

\bibpunct{(}{)}{;}{a}{}{,}

\title{Physics-guided probabilistic modeling of extreme precipitation under climate change}

\authors{Evan Kodra\correspondingauthor{risQ, Inc., 55 Magazine St 6B, Cambridge, MA, 02139}, Singdhansu Chatterjee\thanks{Current affiliation: Department of Statistics, University of Minnesota, Minnesota, MN, 55455; Secondary affiliation: risQ, Inc., Cambridge, MA}, Stone Chen \thanks{Current affiliation: risQ, Inc., 55 Magazine St 6B, Cambridge, MA, 02139} and Auroop R. Ganguly\thanks{Current affiliation: Department of Civil and Environmental Engineering, Northeastern University, Boston, MA, 02115; Secondary affiliation: risQ, Inc., Cambridge, MA}}

\affiliation{risQ, Inc., Cambridge, MA}

\email{evan.kodra@risQ.io}

\abstract{Earth System Models (ESMs) are the state of the art for projecting the effects of climate change. However, longstanding uncertainties in their ability to simulate regional and local precipitation extremes and related processes inhibit decision making. Stakeholders would be best supported by probabilistic projections of changes in extreme precipitation at relevant space-time scales. Here we propose an empirical Bayesian model that extends an existing skill and consensus based weighting framework and test the hypothesis that nontrivial, physics-guided measures of ESM skill can help produce reliable probabilistic characterization of climate extremes. Specifically, the model leverages knowledge of physical relationships between temperature, atmospheric moisture capacity, and extreme precipitation intensity to iteratively weight and combine ESMs and estimate probability distributions of return levels. Out-of-sample validation shows evidence that the Bayesian model is a sound method for deriving reliable probabilistic projections. Beyond precipitation extremes, the framework may be a basis for a generic, physics-guided approach to modeling probability distributions of climate variables in general, extremes or otherwise.}

\begin{document}

%% Necessary!
\maketitle

%%%%%%%%%%%%%%%%%%%%%%%%%%%%%%%%%%%%%%%%%%%%%%%%%%%%%%%%%%%%%%%%%%%%%
% MAIN BODY OF PAPER
%%%%%%%%%%%%%%%%%%%%%%%%%%%%%%%%%%%%%%%%%%%%%%%%%%%%%%%%%%%%%%%%%%%%%
%
\section{Introduction}\label{introduction}
Stakeholder-oriented synthesis reports at all scales, from global \citep{pachauri2014climate} to local \citep{climateReadyBoston2016} rely heavily on ESMs, the principal tools for projecting the effects of climate change. Literature \citep{katz2013uncertainty} and our interactions with stakeholders \citep{kodra2012nrc, ganguly2015climate, climateReadyBoston2016} support the need for probabilistic projections of climate change. However, ESMs do not provide probabilistic projections directly \citep{tebaldiknutti2007, knutti2010challenges}. Probabilistic climate projections can serve as tools for designing structures \citep{mailhot2007assessment, mirhosseini2013impact, shrestha2017developing} and potentially for pricing financial risk mitigation and transfer instruments like short term insurance, reinsurance, catastrophe bonds, and long term insurance \citep{jaffee2010long, kunreuther2011insuring, maynard2012role}. Probabilistic characterization is especially crucial for extremes at regional and local scales \citep{mailhot2007assessment, mirhosseini2013impact, shrestha2017developing}, and yet not many approaches exist \citep{sunyer2014bayesian}.

While a wealth of literature focuses on the global mean response of climate to greenhouse gases (GHGs), often the largest changes are expected to occur in the tails of the distribution of climate variables \citep{trenberth2012framing}. Statistical attributes of extremes, namely their intensity, duration, and frequency, are changing and are expected to continue to do so under climate change \citep{kao2011intensity, wang2016synoptic}. For example, floods that have been considered 1-in-500 year events (or equivalently stated, a flood event that can be met or exceeded in intensity with a $0.2\%$ in any given year) are occurring at a frequency that suggests their true contemporary likelihood may now be substantially higher than 1-in-500 \citep{kao2011intensity, wang2016synoptic}. Current, near-term, and long-term uncertainties about extremes act as a major inhibitor, for example, to the advent of updated design storm curves \citep{mailhot2007assessment, mirhosseini2013impact, shrestha2017developing} and the development of long term insurance initiatives that could yield a more sustainable paradigm for mitigating the effect of changing hazards \citep{jaffee2010long, kunreuther2011insuring, maynard2012role}.

\section{Background}\label{background}
\subsection{Skill, consensus, and physics-guided climate model weighting} \label{skillconsensus}
Two principal approaches to probabilistic climate modeling exist. The first utilizes Perturbed Physics Ensembles \citep{murphy2004quantification, stainforth2005uncertainty}, where one or more ESMs are run many (e.g., thousands) of times with different parameters. Among other issues \citep{tebaldiknutti2007, knutti2010challenges}, the approach is often not practical as it requires massive computational resources. The other approach involves exploiting archived ensembles of ESM runs to estimate probability distributions of climate change. Although several variations have been proposed, perhaps the most well-known and developed is skill- and consensus-based weighting, wherein ESMs in an ensemble are weighted based on their ability to replicate historical climate observations -- \textbf{skill} -- and on their agreement with their peers about the future -- \textbf{consensus} \cite{giorgi2002calculation}. This approach was formalized for regional average temperature and precipitation in a Bayesian framework soon after \citep{tebaldi2004regional, tebaldi2005quantifying}. It was then extended in several studies to accommodate bivariate relationships between averages of climate variables \citep{tebaldi2009joint} and to support efficient probabilistic modeling across multiple geographic regions simultaneously \citep{smith2009bayesian}. To date, most of these studies have only supported averages of climate variables. An exception is a recent study that applies this framework to precipitation extremes \citep{sunyer2014bayesian}. Specifically, it applies a modified version of the framework to the $95^{th}$ percentile of precipitation depth on wet days.

Our current study borrows a large portion of the ideas from the skill- and consensus-based framework \citep{tebaldi2004regional, tebaldi2005quantifying, smith2009bayesian, tebaldi2009joint, sunyer2014bayesian} and extends it for building probability distributions of precipitation extremes in a more generalized fashion, allowing for the guidance of known physics through covariance with other climate variables. Our proposed model is an empirical Bayesian one; from here forward, for brevity, we will refer to it as a Bayesian model. 

Literature has pointed out the difficulty of measuring the ``skill" of an ESM \citep{tebaldiknutti2007, knutti2010challenges, knutti2010end, weigel2010risks}, despite a multitude of attempts to do so \citep{gleckler2008performance, pierce2009selecting, santer2009incorporating}. Furthermore, in many cases common skill metrics such as root mean squared error \citep{gleckler2008performance} tend to not lead to systematic differences in terms of model projections \citep{pierce2009selecting, santer2009incorporating}; that is, a ``better model" often does not say anything different about the future than a ``bad" model. Several notable studies, however, suggest that skill metrics designed to capture whether an ESM is simulating a non-trivial physical process can lead to clearer insights about anthropogenic attribution \citep{santer2013human} or reduced future uncertainty \citep{hall2006using, boe2009september, fasullo2012less}. From this, we can synthesize a hypothesis that non-trivial, physics-guided measures of skill may be more useful indicators of ESM reliability. This hypothesis is tested formally via the Bayesian model proposed in the current study, using precipitation extremes as a case.

\subsection{Physics of precipitation extremes} \label{physics}
Precipitation extremes are in many cases expected to increase in intensity, duration, and/or frequency as a function of climate change given theory \citep{o2009physical, sugiyama2010precipitation, o2012sensitivity}, evidence from observations \citep{min2011human}, and ESM projections \citep{pall2007testing, kao2011intensity}. At a global scale this can be explained by the Clausius Clapeyron (CC) equation \citep{pall2007testing, trenberth2011changes}, which shows that under ideal conditions, atmospheric moisture capacity increases in a warming climate.

The August-Roche-Magnus formula \citep{lawrence2005relationship} provides an empirically derived approximation in ideal conditions (between -40 and 50 degrees Celsius and over a plane surface of water):
\begin{equation}\label{cc}
e_s(T) = 6.1094 \exp{(\frac{17.626T}{T+243.04})}
\end{equation}
where $e_s$ is saturation vapor pressure (i.e., atmospheric moisture holding capacity) and $T$ is temperature in Celsius. Moisture condenses to precipitable water when atmospheric moisture holding capacity is reached. 

Since global total annual precipitation is not expected to change significantly under climate change, this in aggregate implies a shift in the distribution of precipitation. Specifically, larger $e_s(T)$ values imply longer duration between condensation and thus precipitation events. When heavy precipitation events do occur, they are expected to increase in intensity owing to increased atmospheric moisture content. Ultimately then, in aggregate, increasing temperatures under climate change translates to increased capacity for drought risk with simultaneous increased potential for extreme precipitation and flood risk \citep{pall2011anthropogenic}. At a global average scale, it has been estimated that atmospheric moisture capacity increases by 7$\%$ per degree Celsius \citep{trenberth2011changes}. 

However, the intensity of extreme precipitation has been shown to depend not only on atmospheric moisture capacity dictated by average local climatological temperature. Precipitation occurrence and intensity is also a function of the local temperature anomaly and upward vertical wind velocity at the time of the precipitation event \citep{o2009physical, sugiyama2010precipitation,pfahl2017understanding}. Though more relevant in tropical latitudes, the cloud physics that drive the formation of convective precipitation are still relatively poorly simulated and are currently parameterized in ESMs \citep{knutson2004impact, schiermeier2010real, schiermeier2015physicists}. The rate of change of extreme precipitation intensity per unit increase in temperature varies significantly according to the regional importance of each of these other driving factors \citep{pfahl2017understanding}. The relationship between regional extreme precipitation and temperature has also been shown to be more complex and nonlinear in many cases as well, potentially owing to moisture limitations at very high temperatures \citep{wang2017}. 

Generally it would be difficult to assess an ESM's ability to simulate the dynamical processes (upward vertical wind velocities) that partially drive extreme precipitation since observational data for those processes are usually not even available. In contrast, in many regions of the world, high quality observations for both temperature and precipitation do exist. Hence, in this study, we leverage this knowledge with the following hypothesis: a skillful ESM should be able to successfully replicate not only the observed marginal distribution of extreme precipitation but also its observed dependence (or lack thereof) on contemporaneous air temperature at a regional scale. The complexity of the relationship between air temperature and extreme precipitation \citep{wang2017} as well as the relative regional dominance of dynamical processes \citep{pfahl2017understanding} inhibits straightforward CC based extrapolation. This further supports the potential utility in modeling the relationship between temperature and extreme precipitation at a regional scale. 

\section{Bayesian Model} \label{bayesianmodel}

\subsection{Data and Preprocessing} \label{preprocessing}
An ensemble of 15 ESMs from the Coupled Model Intercomparison Project Phase 5 (CMIP5) archive is used in this study. For the years 1950-1999, historical ESM runs are used. For the years 2065-2089, runs from the greenhouse gas scenario RCP8.5 are used. The model presented shortly is run for all 18 continental U.S. Hydrologic Unit 2 (HU2) watersheds provided by the United States Geological Survey's Watershed Boundary Dataset (USGS WBD) \citep{berelson2004mapping}. The Appendix provides metadata on the ESMs and watersheds used.

USGS WBD HU2 shape files are used to identify grid cells that belong to each watershed. In each watershed and for each ESM, preprocessing is conducted as follows. For each month and year, daily total precipitation block maxima are extracted for each grid cell and then averaged over all grid cells within the watershed. For each month separately, those block maxima are then sorted in ascending order and treated as return levels. We sort the block maxima rather than examine them in their original temporal order. We do this because ESMs are not necessarily expected to be in phase with observations or with other ESMs in terms of cycles of climate variability like El Ni\~no/Southern Oscillation or the Atlantic Multidecadal Oscillation \citep{kodra2012evaluation}. Rather, ESMs are designed to simulate the \textit{statistics} of weather, or the \textit{climatology}, in the neighborhood of a given year. A more meaningful way to compare the statistics is to examine ESMs and observations in terms of a climatological window, e.g., 1975-1999. Sorting block maxima in ascending order helps alleviate the fact that ESMs are likely to be out of phase with each other and observations. This idea has been utilized in statistical downscaling; with so called asynchronous regression approaches, the order statistics of observations are regressed on the order statistics of an ESM to create transfer functions that can be carried forward to future ESM simulations \citep{stoner2013asynchronous}. Reordering block maxima does assume that there is no serial correlation between subsequent years and that they are stationary. These are typical assumptions made in extreme value modeling situations and are usually reasonable in climate research if block maxima are far enough apart and can be treated as approximately independent \citep{coles_modelling_1991, kharin2007changes, kodra2014asymmetry}. We further examine these assumptions in the Appendix. Surface (6-meter) air temperature averaged over the same days as the block maxima are extracted and re-sorted according to precipitation ordering, as well. Note that temperature is sorted not in ascending order of temperature but of precipitation. Thus, temperature is not necessarily increasing or decreasing once re-ordered.

Observational precipitation maxima and surface air temperature are extracted from a higher resolution ($\frac{1}{16}$) degree gridded observational data product \citep{livneh2013long} for the years 1950-1999 and are preprocessed in the same manner as the ESMs.

We denote $P$ as return levels of precipitation and $T$ as temperature averaged over the same day in the same location. The subscript $k$ indexes observational datasets (there is only one observational dataset used in this study, but the Bayesian model allows for more than one); $m \in [1,...M=12]$ indexes season (calendar month in this study), $q \in [1,...Q=25]$ indexes the ranks of the return levels from smallest to largest from a historical climatology, $q' \in [1,...Q'=25]$ the same but for the future climatology, and $j$ indexes ESM datasets. 

Let $Z_{j, m}=\log(P_{j, m,q=1})$, i.e., for any ESM dataset $j$ (or $k$ for observations), the smallest value of the precipitation return levels is transformed with a natural log. Then, for larger return levels $q \in [2,...Q]$, we let $U_{j, m,q}=\log(P_{j, m,q} - P_{j, m, q-1})$. In short, there is a need to ensure that realizations from the Bayesian model are larger than 0. There is also a need to ensure that realizations $P_{m,q} > P_{m,q-1}$, i.e., that higher order statistics are always larger than lower ones. These transformations will ensure both of these features. The natural log transformation also generally creates data that are also more amenable to Gaussian data models, an assumption checked and discussed in the Appendix.

This preprocessing is done for three separate climatologies: 1950-1974, 1975-1999, and 2065-2089. The use of these three climatologies is summarized in Table \ref{schemes}.

The validation scheme is particularly important for assessing performance of the Bayesian model in terms of accuracy and posterior coverage. In that scheme, 1950-1974 is the historical climatology and 1975-1999 the future climatology (where in this case, 1975-1999 is treated as a ``hold out"). Methodology used for validation is discussed more in section \ref{bayesianmodel}\ref{validation}. For the end of century scheme where observational data is not available in the future time regime, probabilistic prediction is of interest. In this case, 1975-1999 is used as the historical climatology.

\subsection{Data Model} \label{datamodel}
We leverage the Bayesian skill and consensus-based framework discussed earlier \citep{tebaldi2004regional, tebaldi2005quantifying, smith2009bayesian, tebaldi2009joint, sunyer2014bayesian} as the mechanical foundation for our model. Through a Markov Chain Monte Carlo (MCMC) process, ESM projections of return levels are iteratively weighted and averaged according to (1) their skill as measured by their similarity to observational return levels and (2) to a lesser extent, their consensus with projections. Skill is formulated to explicitly evaluate whether the return levels from ESMs depend on temperature in the same way that they do in observations. 

First, the smallest of the return levels are assumed to follow Gaussian models: 

\begin{equation} \label{eq1}
Z_{k,m} \sim N(C_m, (\tau_k\sigma_k)^{-1})
\end{equation}

\begin{equation} \label{eq2}
Z_{j,m} \sim N(C_{m} + CBIAS_{j},  \sigma_j^{-1})
\end{equation} 

\begin{equation} \label{eq3}
Z'_{j,m} \sim N(C'_{m} + CBIAS_{j}, (\theta\sigma_j)^{-1})
\end{equation} 

The unknowns $C_{m}$ and $C'_{m}$ are seasonal parameters that can be estimated given that there are multiple models and observational datasets. In practice in this study, we set $C_{m}$ as fixed and estimated from historical data as $C_{m}=\log(P_{k,m,q=1})$. $CBIAS_{j}$ is a bias term for ESM $j$ but is assumed to be constant over time regimes. The parameter $\sigma_{j}$ is a scalar weight for each ESM since we have only $M$ seasons over which to estimate it. Finally, $\theta$ is a future variance scaling parameter that modulates the importance of consensus in the determination of weights and also allows for larger uncertainty in the future climatological regime \citep{ganguly2013computational}.

Similar to models from past studies \citep{tebaldi2004regional, tebaldi2005quantifying, smith2009bayesian}, the weight parameter $\sigma_{k}$ is estimated from observational data as:

\begin{equation}
\sigma_{k} = (\frac{\sum_{m=1}^{M}(Z_{k,m} - \bar{Z}_{k,m})^2} {M-1})^{-1}
\end{equation}
i.e., the inverse of the sample variance of the smallest block maxima ($q=1$) over all $M$ seasons.

Next, we define the data model for values of $U_{w, m,q}$ and $U'_{w, m,q'}$, for $q \in [2,...Q]$ and $q' \in [2,3,...Q']$.

\begin{equation} \label{eq4}
U_{k,m,q} \sim N(\gamma_{m, q} + \phi_m\delta_{k,m,q}, (\tau_k\epsilon_{k,q})^{-1}) 
\end{equation}

\begin{equation} \label{eq5}
\begin{split}
U_{j,m,q} \sim N(\gamma_{m, q} + \alpha_{j,m} + \phi_m\delta_{j,m,q}, (\epsilon_{j,q})^{-1})
\end{split}
\end{equation} 

\begin{equation} \label{eq6}
\begin{split}
U'_{j,m,q'} \sim N(\gamma'_{m, q'} + \alpha_{j,m} + \phi'_m\delta'_{j,m,q'}, (\beta'_{m,q'}\epsilon_{j,q})^{-1})
\end{split}
\end{equation} 

In practice, we estimate $\gamma_{m,q}$ as fixed using historical data as $\gamma_{m,q}=\log(P_{k,m,q} - P_{k,m,q-1})$. Since both $C_{m}$ and $\gamma_{m,q}$ are estimated as fixed from historical data in the vein of past studies \citep{tebaldi2004regional, tebaldi2005quantifying, smith2009bayesian}, effectively observed historical precipitation maxima are not random variables and are assumed to be truth. Using metrics in section \ref{bayesianmodel}\ref{validation}, we found that the model performs similarly but slightly better overall with $C_{m}$ and $\gamma_{m,q}$ as fixed versus as random variables. The treatment of $C_{m}$, $\sigma_j$, $\gamma_{m,q}$, and $\epsilon_{j,k}$ as fixed and estimated from data is the empirical aspect of the Bayesian model proposed here. 

The variable $\delta_{j,m,q}=T_{j,m,q} - T_{j,m,q-1}$ is an abbreviation for any ESM $j$ (or observational dataset $k$). Again similar to past related studies \citep{tebaldi2004regional, tebaldi2005quantifying, smith2009bayesian}, we fix $\epsilon_{k,q}^{-1}$ as follows: first, separately for each season $m$, we fit a simple linear regression $U_{k,m=m,q\in [2,...Q]} \sim \delta_{k,m=m,q\in [2,...Q]}$. We save the residuals from each of these regressions. Then, for each order statistic $q$, we calculate the sample variance of the residuals for using $q$ all seasons $m \in [1, ... M]$ (12 data points in this case, where $M=12$). These calculations assume that pooling information across seasons is an approach that yields reasonable information on variability. 

With Eqs. \ref{eq4} - \ref{eq6}, we are essentially assuming that the logarithm of the differential between any pair of subsequent order statistics, which is a sample quantity, is Gaussian with the mean being the population equivalent. As such, the model does not suggest that extremes themselves are Gaussian, it merely says sample versions are normally distributed around true population quantities. This Gaussian assumption of the $U_{k,m,q}$ statistics is examined more in the Appendix.

We hypothesize that observational climate data should indeed more heavily influence historical true climate than historical ESM runs, but it is also important to keep in mind that observations themselves are potentially noisy realizations of the truth. The parameter $\tau_k$ is set as a fixed constant to scale the weight parameters associated with observational data, $\sigma_k$ and $\epsilon_{k,q}$. If $\tau_{k}=1$, this effectively means that the weights $\sigma_k$ and $\epsilon_{k,q}$ are simply the inverse variances as described above. However, the Bayesian model does not necessarily treat observations as ground truth in the way a supervised learning problem would. The parameter $\tau_k$ lets us manually scale the weight of observational climate data to behave more like ground truth, which in turn influences values of unknown parameters in a manner similar in spirit to a supervised learning problem. We explore the sensitivity of model results to choice of $\tau_{k}$ in the Appendix but ultimately settle on $\tau_k = 100$. 

The full joint distribution of all 10 unknown parameters: $C'_m$, $CBIAS_j$, $\sigma_j$,  $\theta$,  $\phi_m$, $\alpha_{j,m}$, $\epsilon_{j,q}$, $\gamma'_{m, q'}$, $\phi'_m$, and $\beta_{m, q'}$ is not of an analytically known form. Similar to past studies \citep{tebaldi2004regional, tebaldi2005quantifying, smith2009bayesian}, we choose conjugate prior distributions for each unknown that lead to known full conditional posterior distributions. In other words, for example $\gamma'_{m, q'}$ (or any other of the unknowns) follows a posterior distribution of  known form (in this case, Gaussian) assuming all other unknown values are known. All unknowns are updated in a Gibbs sampler variant of a Markov Chain Monte Carlo (MCMC) simulation. The Appendix provides full details on the prior parameters, sensitivity tests for key prior parameters, and the full conditional posterior distribution for all unknowns. The Appendix also provides complete information on MCMC simulations and associated diagnostics.

\subsection{Bayesian Model Validation} \label{validation}
Validation of the Bayesian model is a crucial component of assessing its utility. Of course, unlike weather forecasting, true validation over future climatologies is impossible in the immediate term given the lead times of interest. We validate the model using a training-holdout scheme similar to conventional predictive modeling. As mentioned in section \ref{bayesianmodel}\ref{preprocessing}, we do this in each region using 1950-1974 as the ``training" and 1975-1999 as the ``validation" climatologies, respectively. 

Once posterior samples are converted back to original units (see Appendix), we examine Bayesian model accuracy, posterior coverage, posterior upper coverage, and posterior width, all as compared to the original ensemble of ESMs.  Accuracy is measured via Root Mean Squared Error (RMSE) of the posterior mean (or original ensemble mean) return levels with reference to held out observations. Coverage is measured as the percentage of held out observations that fall within upper and lower bounds of the posterior distribution (or original ensemble bounds). Similarly, upper coverage is measured as the percentage of held out observations that fall below the upper bounds of the posterior distribution (or original ensemble upper bounds). Width is measured as the average distance from the lower to upper bounds of the posterior distribution (or original ensemble bounds). Width is examined since it is trivial for extraordinarily wide bounds to exhibit high coverage of held out observations. A superior Bayesian model would exhibit higher accuracy than the original ensemble and appropriate coverage (a 99\% credible interval should cover $99\%$ of held out observations). If the original ensemble is inappropriately narrow, the ideal Bayesian posterior bounds would be wider. Likewise, if the original ensemble bounds are inappropriately wide, the ideal Bayesian posterior bounds would be narrower.

In addition, we also compare posterior projected \textit{changes} in return levels as compared to those projected changes obtained directly from the original ensemble of ESMs. For this final measure, where the original ensemble performs well with reference to held out observations, the ideal Bayesian model should exhibit similar projected changes. In cases where the original ensemble performs poorly against held out observations, the ideal Bayesian model might deviate in terms of projected changes. The Appendix provides complete details on accuracy, coverage, width, and return level change calculations and analysis.

\section{Results} \label{results}
\subsection{Validation Results} \label{validation_results}
\begin{figure}[h]
	\includegraphics[width=\linewidth]{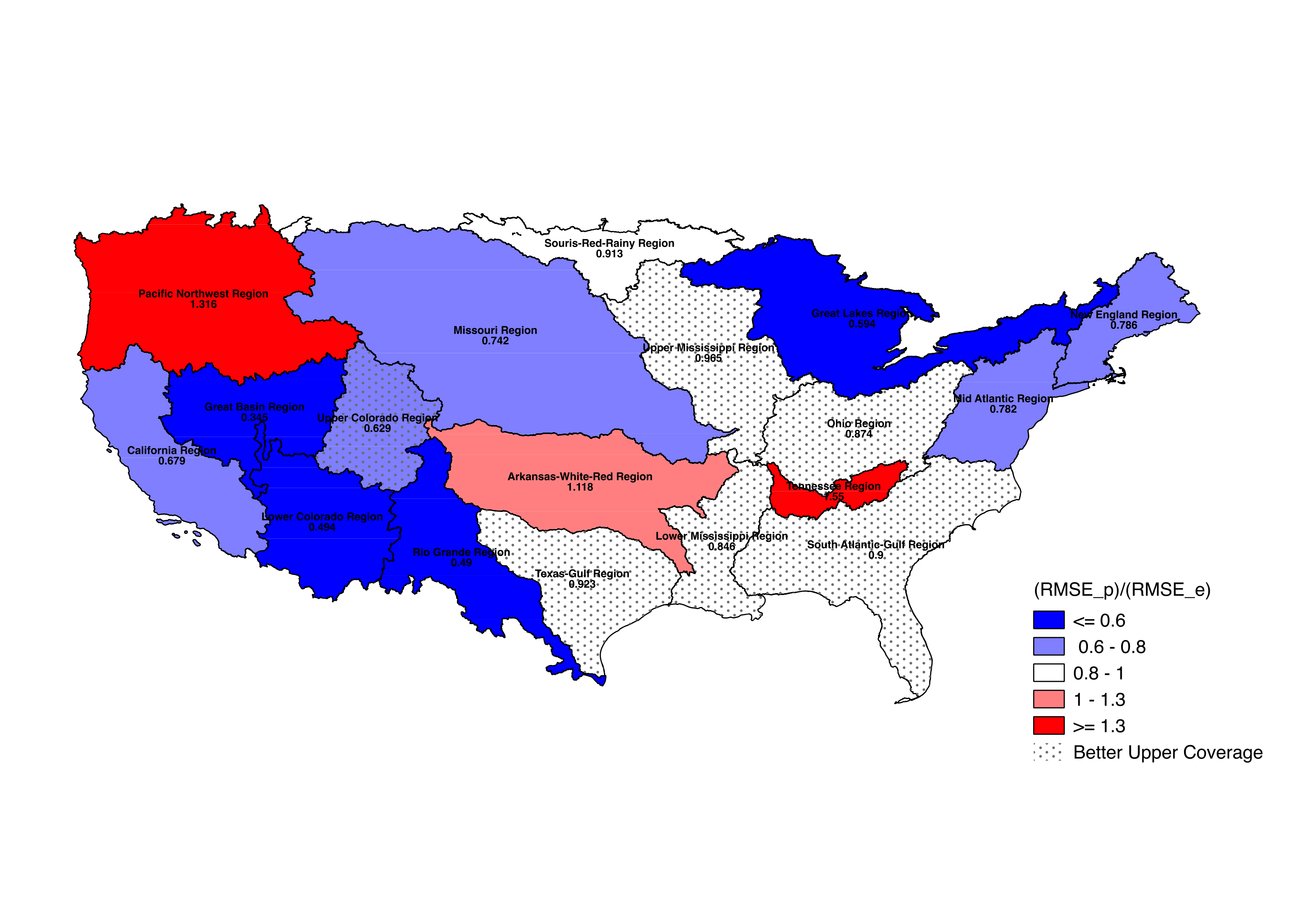}
	\caption{The performance of the Bayesian model is compared to using the raw ensemble in terms of out-of-sample accuracy and predictive coverage across 18 watersheds that comprise the continental U.S. Coloring represents accuracy of the posterior relative to using an ensemble average approach, measured as $\frac{RMSE_{p}}{RMSE_{e}}$. Accuracy is higher in 15 out of 18 watersheds. In 15 of 18 watersheds, using a $99\%$ credible interval, posterior coverage is larger than or equal to than ensemble coverage in all watersheds, where coverage ranges from 0 to 1 (not depicted). The three regions where posterior coverage is smaller than that of the original ensemble are the Tennessee, Pacific Northwest, and California watersheds. Stippling here indicates watersheds where upper posterior coverage is larger than or equal to ensemble upper coverage; upper coverage is equivalent or improved in only 6 out of 18 watersheds using the same $99\%$ credible interval. Watersheds are labeled by name and their respective $\frac{RMSE_{p}}{RMSE_{e}}$ values. Section \ref{bayesianmodel}\ref{validation} and the Appendix provide more on definitions and calculations of RMSE, coverage, and upper coverage.} 
	\label{HU2ValidationPerformanceMap}
\end{figure}

\begin{figure}[h]
\centering
	\includegraphics[width=\linewidth]{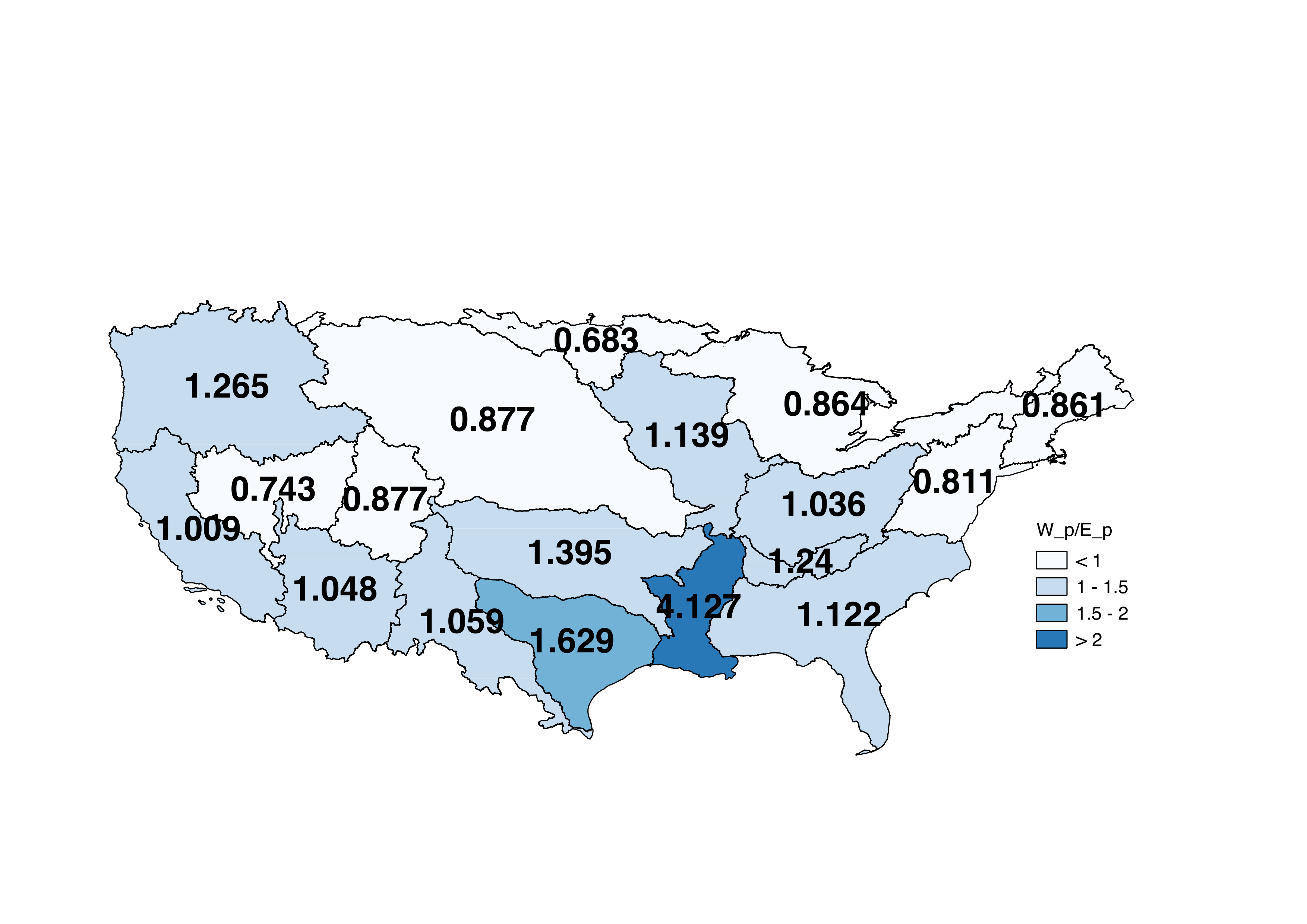}
	\caption{The ratio $\frac{W_p}{W_e}$ (posterior width over original ensemble width) is displayed for all 18 watersheds. Watersheds are also labeled by $\frac{W_p}{W_e}$. The posterior width $W_p$ is calculated from a $99\%$ credible interval while $W_e$ uses the full ensemble bounds (see section \ref{bayesianmodel}\ref{validation}). The blue color palette was chosen for this map as a neutral gradient, since the optimal values of $W_{p}$ depend on the context of the accuracy and posterior coverage attributes as shown in Figure \ref{HU2ValidationPerformanceMap}. Section \ref{bayesianmodel}\ref{validation} and the Appendix provide more on the definition and calculation of width.} 
	\label{HU2ValidationWidthMap}
\end{figure}

Figures \ref{HU2ValidationPerformanceMap} - \ref{HU2ValidationWidthMap} synthesize validation scheme results across the 18 USGS HU2 watersheds that comprise the continental United States. Out-of-sample RMSE-based accuracy, posterior coverage, posterior upper coverage (Figure \ref{HU2ValidationPerformanceMap}), and posterior distribution width (Figure \ref{HU2ValidationWidthMap}) are characterized for each watershed. In the majority of watersheds (15 of 18), the Bayesian model outperforms the equal weighted ensemble average relative to held out observations from 1975-1999 in terms of RMSE-based accuracy. For 15 of 18 watersheds (but not the same 15), the Bayesian model equals or outperforms the raw ensemble upper and lower bounds in terms of posterior coverage when using a $99\%$ posterior credible interval. However, the Bayesian model exhibits better upper coverage in only 6 of 18 watersheds. Also in terms of a $99\%$ credible interval, the posterior distribution is on average wider than ensemble bounds in 11 of 18 watersheds.

\begin{figure}[h]
	\includegraphics[width=\linewidth]{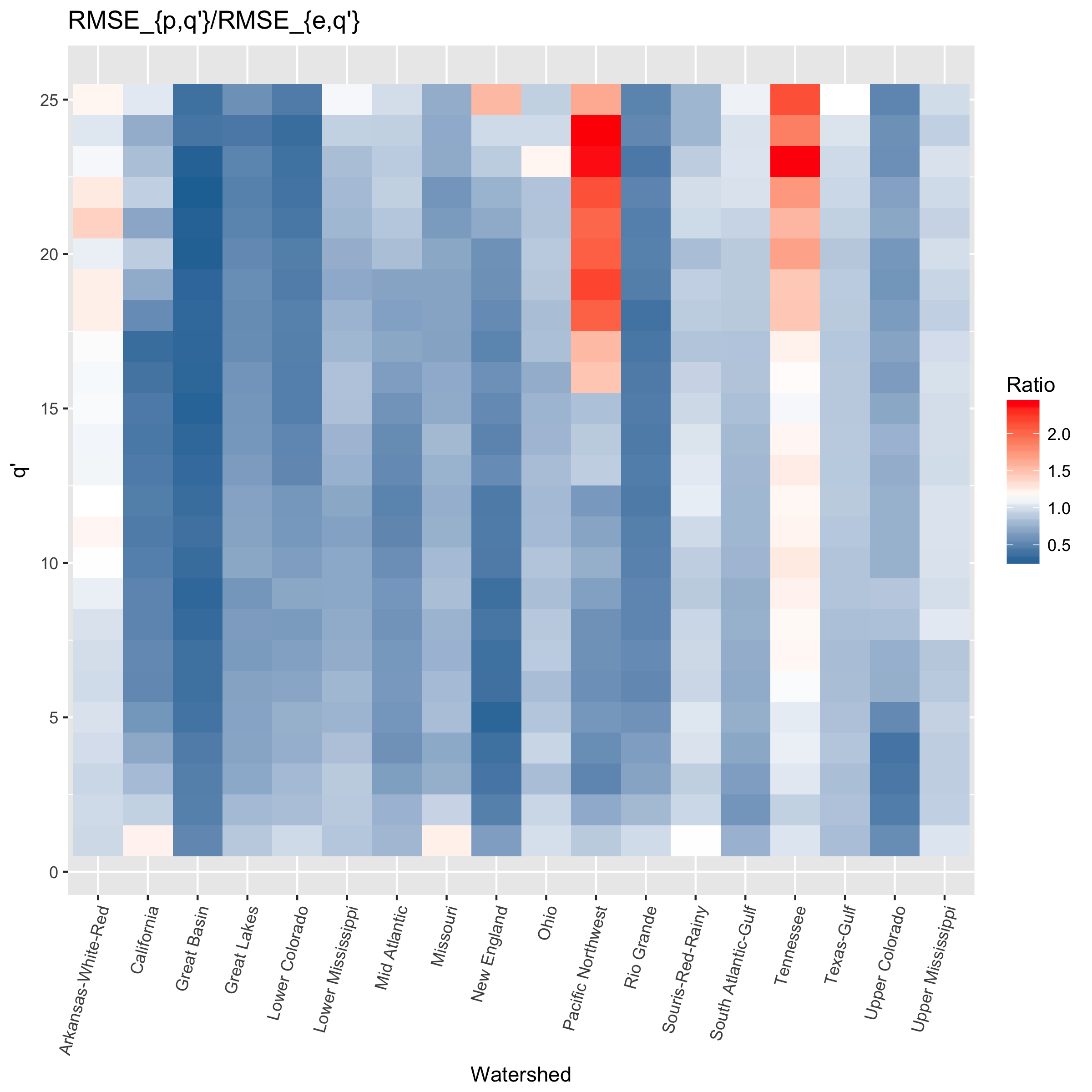}
	\caption{Validation climatology accuracy of the Bayesian model versus the original ensemble is assessed marginally for each return level $q' \in [1,2,...25]$ and each watershed. Values of $\frac{RMSE_{p, q'}}{RMSE_{e, q'}}$ (posterior RMSE over ensemble RMSE, see Appendix) are shown with a heatmap. Blue colored cells are cases where the Bayesian model is more accurate than the original ensemble, and vice versa for red cells.} 
	\label{quantile_rmse_heatmap}
\end{figure}

\begin{figure}[h]
	\includegraphics[width=\linewidth]{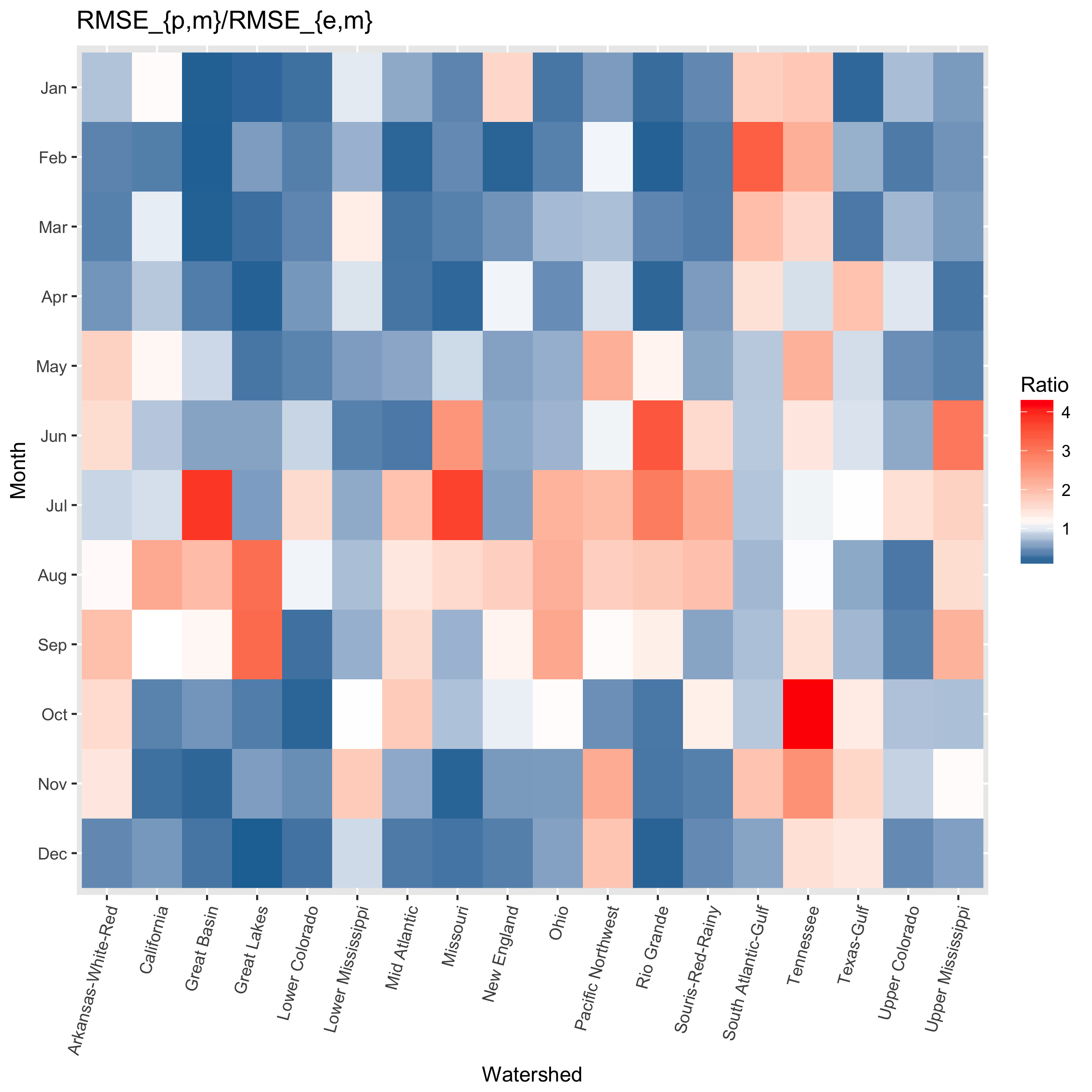}
	\caption{The same is shown as Figure \ref{quantile_rmse_heatmap} but marginally for every month and watershed. Colors correspond to values of $\frac{RMSE_{p, m}}{RMSE_{e, m}}$ (posterior RMSE over ensemble RMSE, see Appendix) for the validation climatology.} 
	\label{seasonal_rmse_heatmap}
\end{figure}

Figure \ref{quantile_rmse_heatmap} shows the accuracy of the Bayesian model compared to the original ensemble marginally for every return period $q' \in [1,2,...25]$. The ratio $\frac{RMSE_{p,q'}}{RMSE_{e,q'}}$ is plotted as a heatmap for every watershed and return period $q'$, where $RMSE_{p,q'}$ is the RMSE of the posterior for only $q'$ and $RMSE_{e,q'}$ the same but for the original ensemble. The Bayesian model outperforms the ensemble in $\sim82.7\%$ (372 out of 25x18=450) of cases. In several watersheds, most notably Tennessee and the Pacific Northwest, the Bayesian model performs poorly compared to the original ensemble for progressively higher return periods. Figure \ref{seasonal_rmse_heatmap} is the same but marginally for each month using the ratio $\frac{RMSE_{p,m}}{RMSE_{e,m}}$, where $RMSE_{p,m}$ is the RMSE of the posterior for only season $m$ and $RMSE_{e,m}$ the same but for the original ensemble. The Bayesian model outperforms the ensemble in $62.5\%$ (135 out of 12x18=216) of cases. Compared to Figure \ref{quantile_rmse_heatmap} where the gradient in some watersheds obviously changes across $q'$, the pattern is less smooth here. Overall, the Bayesian model tends to be more accurate than the ensemble in non-summer months. In June through August, the ensemble often outperforms the Bayesian model. Combined information from Figures \ref{quantile_rmse_heatmap} and \ref{seasonal_rmse_heatmap} suggests that, specifically, the Bayesian model struggles to perform well for the most intense summer precipitation events. Though interpreting why is not straightforward, among many potential reasons, for example, literature \citep{trenberth2003changing} points out that the diurnal cycle in precipitation is particularly pronounced over the United States in the summer, but that ESMs do a poor job simulating this. As another example, literature \citep{ting1997summertime} has also shown a significant correlation between tropical and North Pacific sea surface temperatures and summer precipitation variability in the United States. Weighting ESMs according to their ability to to capture dependence only on same day temperature, then, might not adequately reflect the physical processes that drive some regions' precipitation in summer months. Bringing additional covariates into the Bayesian model could help improve results in these types of cases.

\begin{figure}[h]
   \centering
    \includegraphics[width=\linewidth]{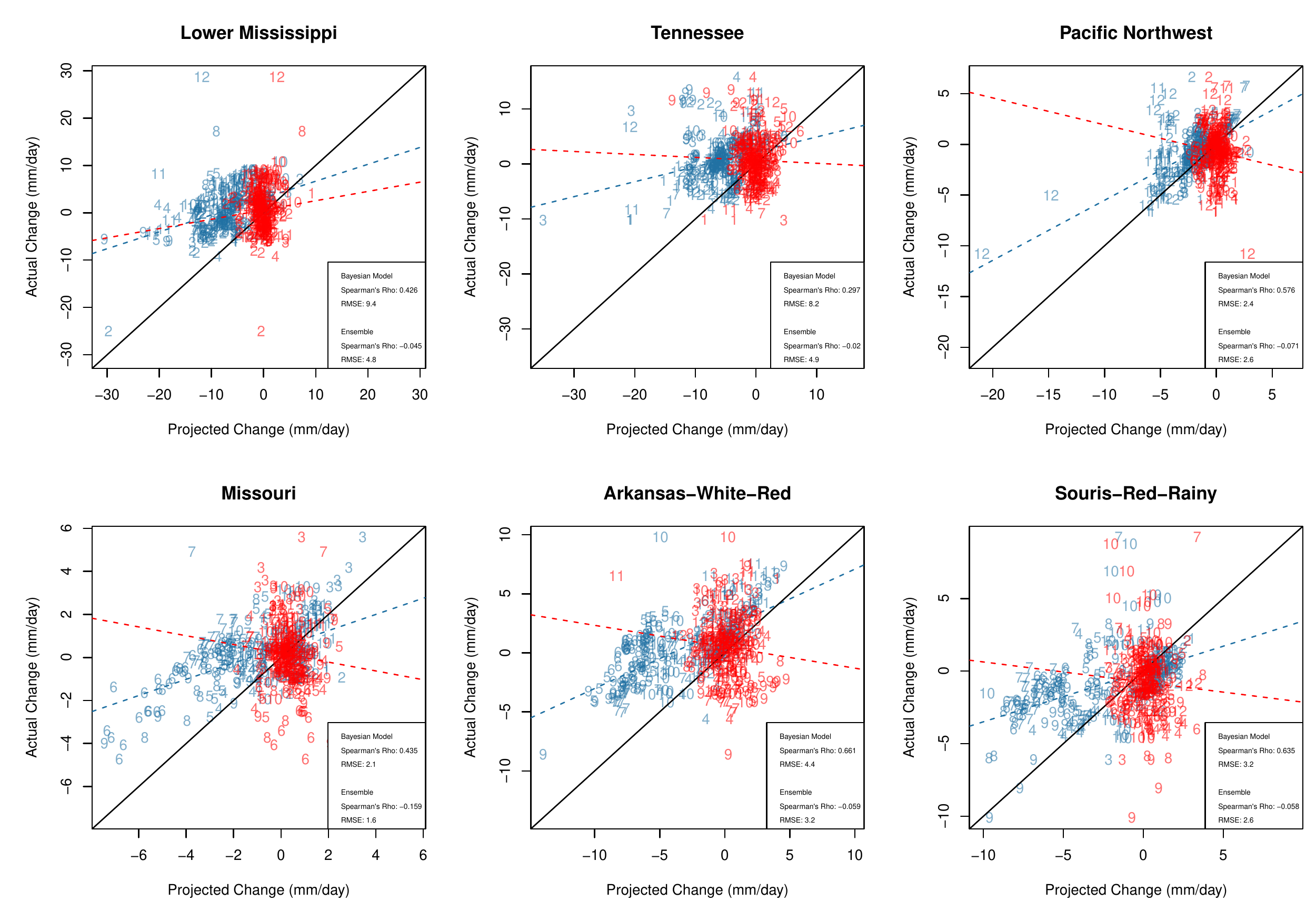} 
    \caption{For 6 watersheds, projected versus observed changes (from 1950-1974 to 1975-1999) in mixed N-year return levels of daily total precipitation depth (mm/day) are shown. Blue numbers correspond to the posterior median projected changes and red to the median of the ensemble. Numbers represent calendar months. A total of 12 months x 25 different return levels = 300 points are shown for both the posterior and ensemble projected changes. The blue line is a least squares line fit between the posterior changes and actual changes, and the red is the same for the original ensemble. The black line is set at $x=y$ for context. }    
    \label{firstsixvalidationchanges}                                                                  
\end{figure}

\begin{figure}[h]
   \centering
\includegraphics[width=\linewidth]{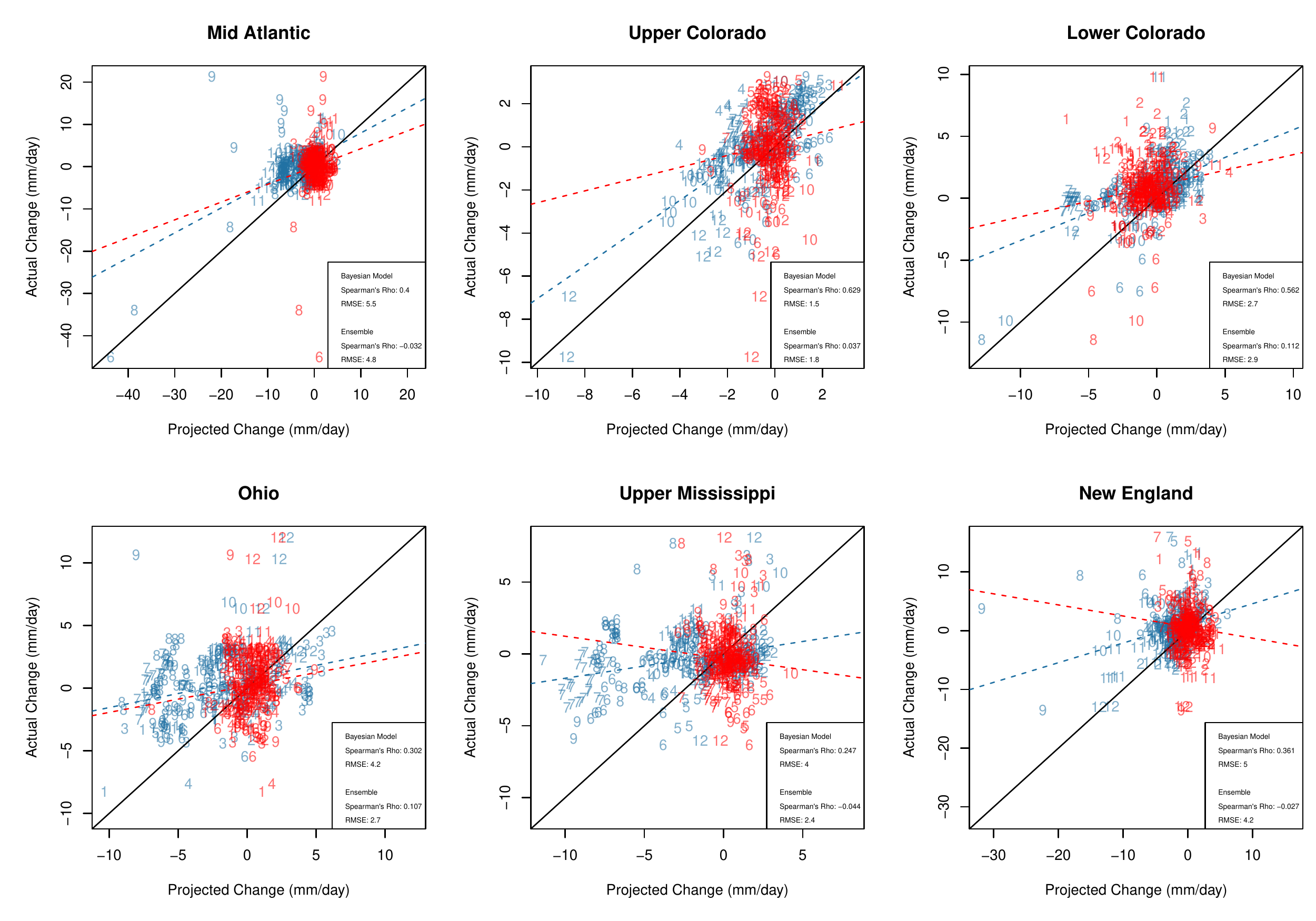} 
\caption{The same as Figure \ref{firstsixvalidationchanges} but for 6 other watersheds.} 
\label{secondsixvalidationchanges}                                                                                       
\end{figure}

\begin{figure}[h]
     \centering
     \includegraphics[width=\linewidth]{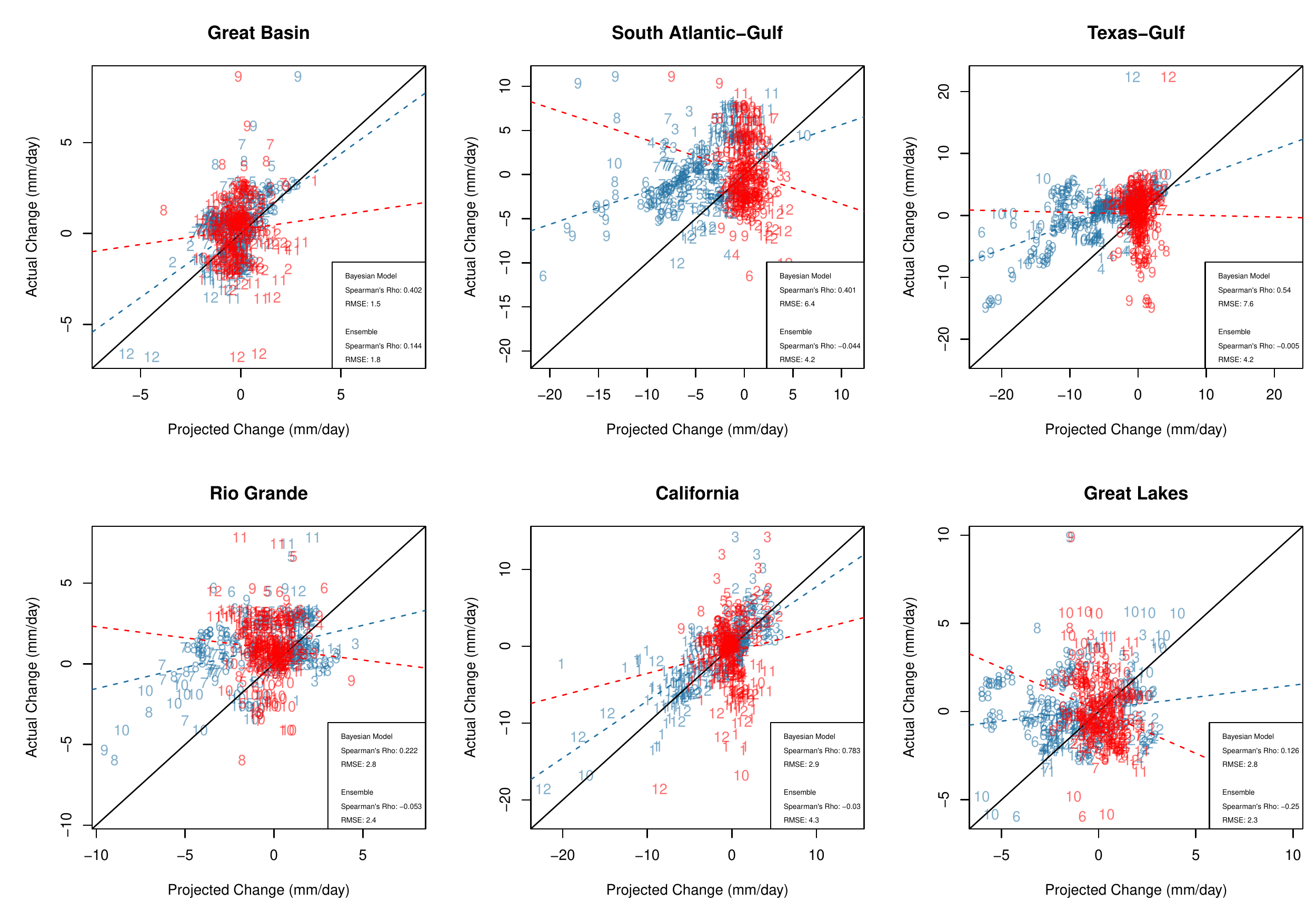}   
\caption{The same as Figure \ref{firstsixvalidationchanges} but for the 6 final watersheds.} 
\label{thirdsixvalidationchanges}                                                                                        
\end{figure}

Figures \ref{firstsixvalidationchanges} - \ref{thirdsixvalidationchanges} measure the ability of the Bayesian model to simulate historical changes in return levels compared to the original ensemble projections. The calculation of these projected changes is further discussed in the Appendix. Projected changes along with actual changes are visualized with scatterplots. Spearman's rank correlation and RMSE are both computed between the projected and actual changes as well. All changes are in $\frac{mm}{day}$ units. The Bayesian model shows a higher correlation with observed changes in return levels across months and return levels than the original ensemble in every watershed. However, in the majority (13/18) of watersheds the original ensemble exhibits a lower RMSE of change. This is because the Bayesian model has a median tendency to underestimate changes in most watersheds; this is evident given that multiple points fall above the diagonal $x=y$ lines. This suggests the possibility that post-model bias correction could produce Bayesian projections that are both correlated with observed changes and more accurate in absolute value relative to the original ensemble. However, validation of that premise is not easily possible without another separate independent holdout climatology; thus, this is left as a hypothesis in the current study.

\subsection{Projections} \label{projections}
\begin{figure}[h]
     \centering
     \includegraphics[width=\linewidth]{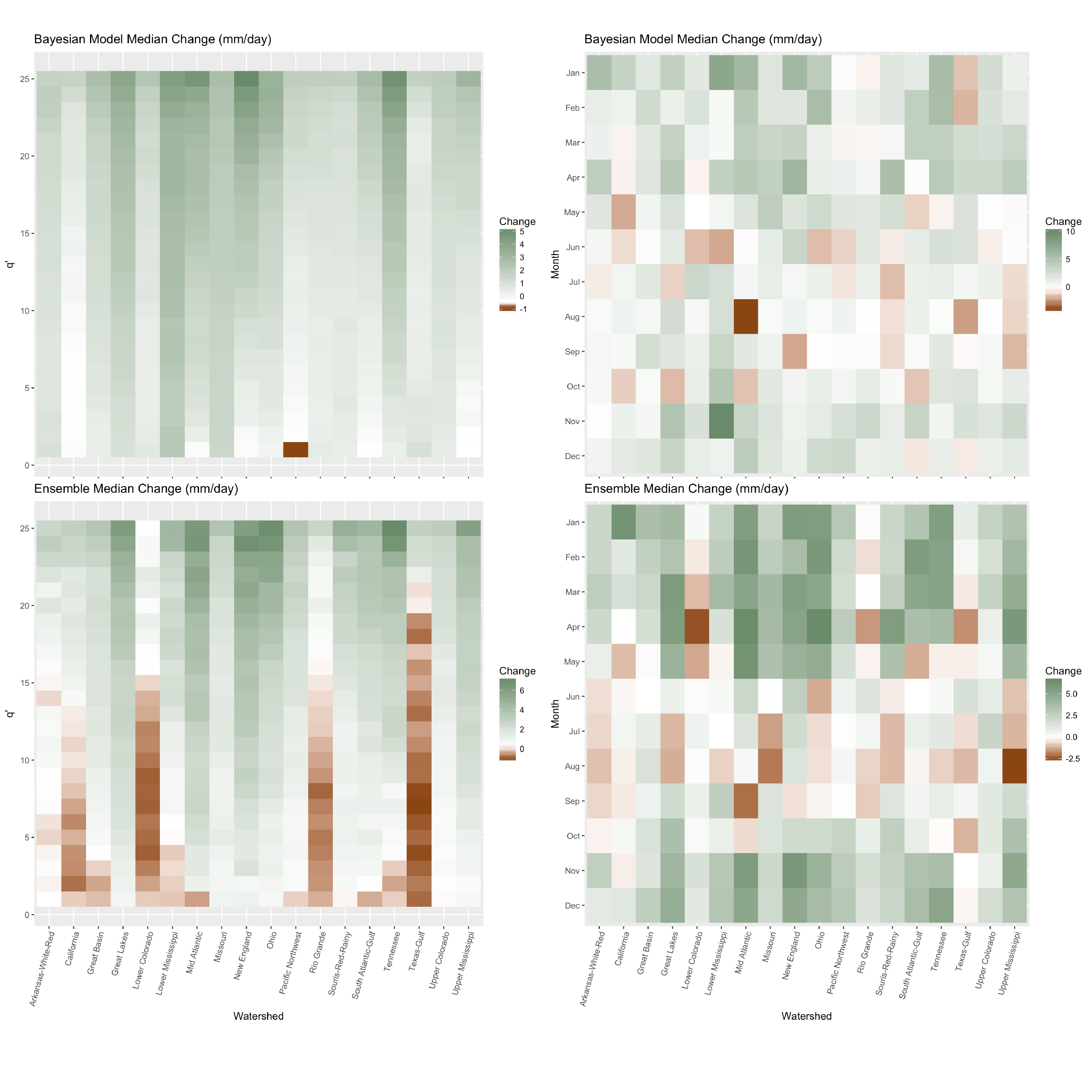}             
\caption{(Top left) Median projected changes are shown for the Bayesian model for each return level, over all months. (Bottom left) The same is shown but for the medians of the original ensemble. (Top right) Median projected changes are shown for the Bayesian model for each season, over all return levels. (Bottom right) The same is shown but for the medians of the original ensemble.} 
\label{heatmaps}                                                                                        
\end{figure}

Figure \ref{heatmaps} provides a comprehensive look at median change projections from both the Bayesian model and the original ensemble, but without information on uncertainty. Changes here are calculated using a slightly different method (see Appendix for detail) than for Figures \ref{firstsixvalidationchanges} - \ref{thirdsixvalidationchanges} since the Bayesian model was shown to typically underestimate change. Heatmaps show medians of posterior projections and the original ensemble for all combinations of return levels and watershed, averaged over all calendar months in each case. Results for both the original ensemble and the Bayesian model in Figure \ref{heatmaps} resemble what could be generally expected under CC scaling in every watershed, where progressively further into the upper tail of the extreme precipitation distribution, intensity increases more (Pall et al. 2007). For the Bayesian model, $\sim 93\%$ (417 of 450) of heatmap cells show increases in return levels, compared to $\sim 84\%$ (376 of 450) for the original ensemble. 

Figures \ref{heatmaps} also shows the same but show median projected changes for each month, averaged over all return levels, for the Bayesian model and original ensemble, respectively. Here, $\sim74\%$ (159 of 216) of cells show an increase for the Bayesian model, whereas $\sim72\%$ (156 of 216) do for the original ensemble. The seasonal pattern of change is similar for both Bayesian and original ensemble projections, with June through September showing more cases of average decrease and the rest of the year showing increases more frequently across the majority of watersheds.

\begin{figure}[h]
     \centering
        \includegraphics[width=\linewidth]{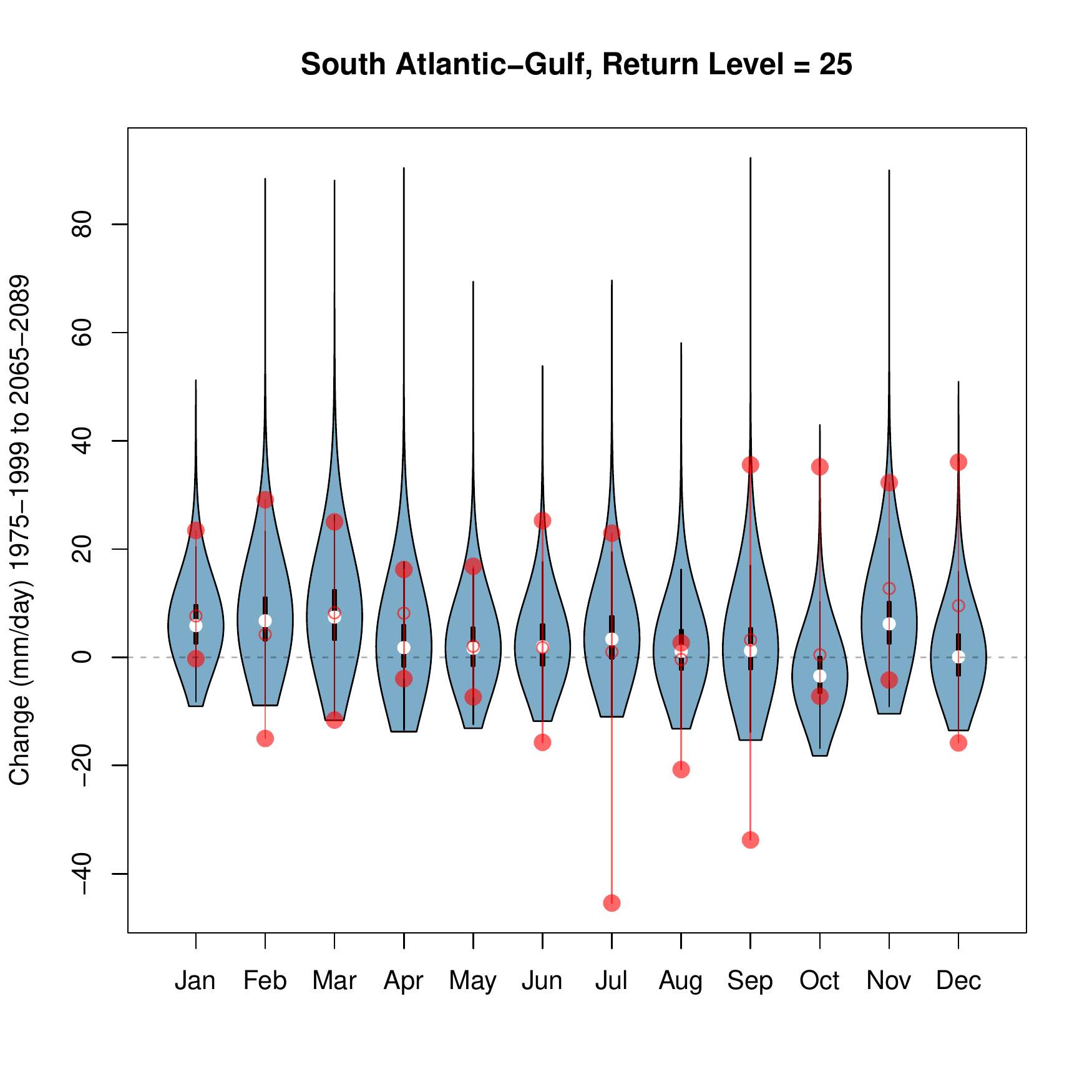}       
\caption{Blue violin plots depict kernel densities of Bayesian probability distributions of projected change (1975-1999 to 2065-2089) in $q'=25$-year return levels in the South Atlantic-Gulf watershed for each month. White dots represent the median of the Bayesian posteriors, and thick and thin black whiskers are lower and upper fences seen in a standard boxplot. Red hollow dots represent the median of the original ensemble projected changes. Red filled dots represent the upper and lower bounds of the original ensemble. } 
\label{south-atlantic-gulf-projected-change-rl25_type2}                                                                                        
\end{figure}

Figure \ref{south-atlantic-gulf-projected-change-rl25_type2} shows the detailed end of century change projections (1975-1999 to 2065-2089) for $q'=25$ year return levels for the South Atlantic-Gulf watershed as an example. Violin plots show a full probability distribution of change relative to the 1975-1999 from the Bayesian model for each calendar month. Median, lower bound, and upper bound change projections from the original ensemble are overlaid for comparison. One notable feature in the Bayesian projections that is absent in the original ensemble is a long upper tail. This lending explicit likelihood to changes that are larger than the original ensemble project. 

Though generally similar on average, the Bayesian change projections have the advantage over the original ensemble in that they provide stakeholders with information on probabilities versus discrete, unweighted projections. 

\subsection{Skill versus Consensus} \label{skillvsconsensus}
\begin{figure}[h]
	\includegraphics[width=\linewidth]{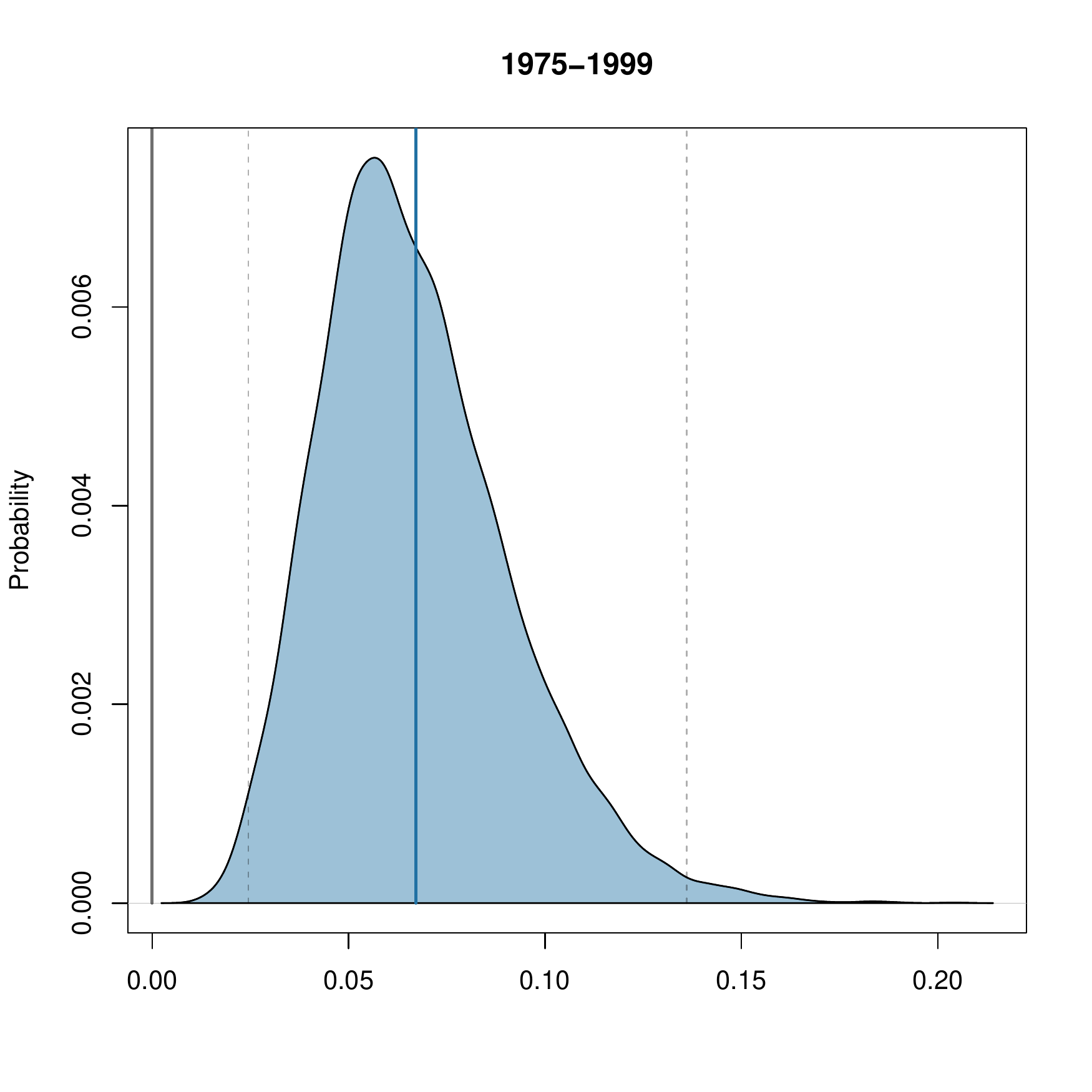}
	\caption{Posterior of $\theta$, a future variance scaling parameter for $q, q'=1$, is shown for the validation scheme model run (1950-1974 as training, 1975-1999 as validation) in the South Atlantic-Gulf watershed. Values are substantially less than 1, meaning that consensus is favored less than skill in weighting ESMs for determining the posteriors of $C'_{m}$.}
	\label{theta}
\end{figure}

\begin{figure}[h]
\centering
        \includegraphics[width=\linewidth]{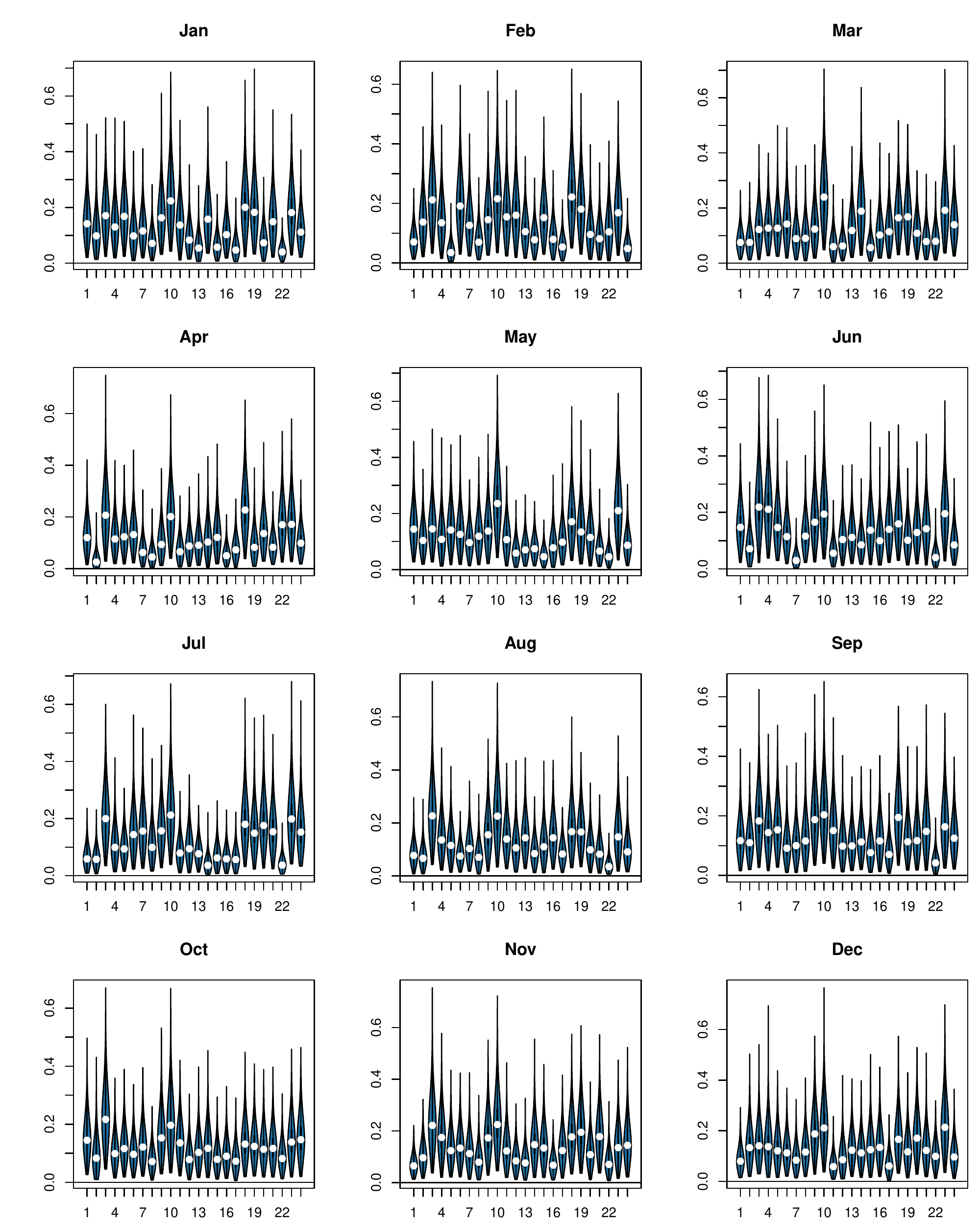} 
      \caption{Posteriors of $\beta'_{m,q'}$ are shown via violin plots for the validation scheme model run in the South Altantic-Gulf watershed. Horizontal axes range from 1 to 24, which map to $q, q' \in [2, ... Q=Q'=25]$.}
    \label{betaprimemqprime}
\end{figure}

Comparatively, previous literature using the skill-consensus framework have suggested the possibility that consensus among ESMs about the future can receive too much emphasis relative to skill \citep{tebaldi2004regional, ganguly2013computational}. In this study, we find that prior parameter choices that tend to produce posteriors that perform well in terms of out-of-sample validation also tend to produce values of $\theta$ and $\beta'_{m,q'}$ that are usually smaller than 1, effectively emphasizing skill over consensus. Figures \ref{theta} and \ref{betaprimemqprime} show posteriors for these parameters for the South Atlantic-Gulf watershed (results are similar for all watersheds). Skill is generally favored over consensus $\sim$ 15-to-1 for $q'=1$ on average according to $\theta$ and $\sim$ 2-to-1 for $q' \in [2,...Q'=25]$ according to $\beta'_{m,q'}$.

The Appendix provides complete posterior results for all other unknown parameters for the South Atlantic-Gulf watershed.

\subsection{Significance of Temperature as a Covariate} \label{tempcovariate}
One of the principal hypotheses of this study is that guiding the statistical architecture of the model with known physics will enhance the results, potentially in a number of ways. In this case the hypothesis centers on the inclusion of same day temperature as a covariate. To test this, we run an experiment with one variant evaluating the model's performance in terms of RMSE performance, posterior coverage, posterior upper coverage, posterior distribution width (all of of $\gamma'_{m,q'}$) while including $\phi_{m}$ and $\phi'_{m}$ as random unknowns (which produces the main results shown in Figures \ref{HU2ValidationPerformanceMap} - \ref{betaprimemqprime}) and another variant where we set $\phi_m=\phi'_m=0$, effectively removing the notion of temperature dependence. We then perform a meta-analysis of the model's performance against the validation regime (see section \ref{bayesianmodel}\ref{validation}) with versus without temperature dependence. 

\begin{figure}[h]
\centering
	\includegraphics[width=\linewidth]{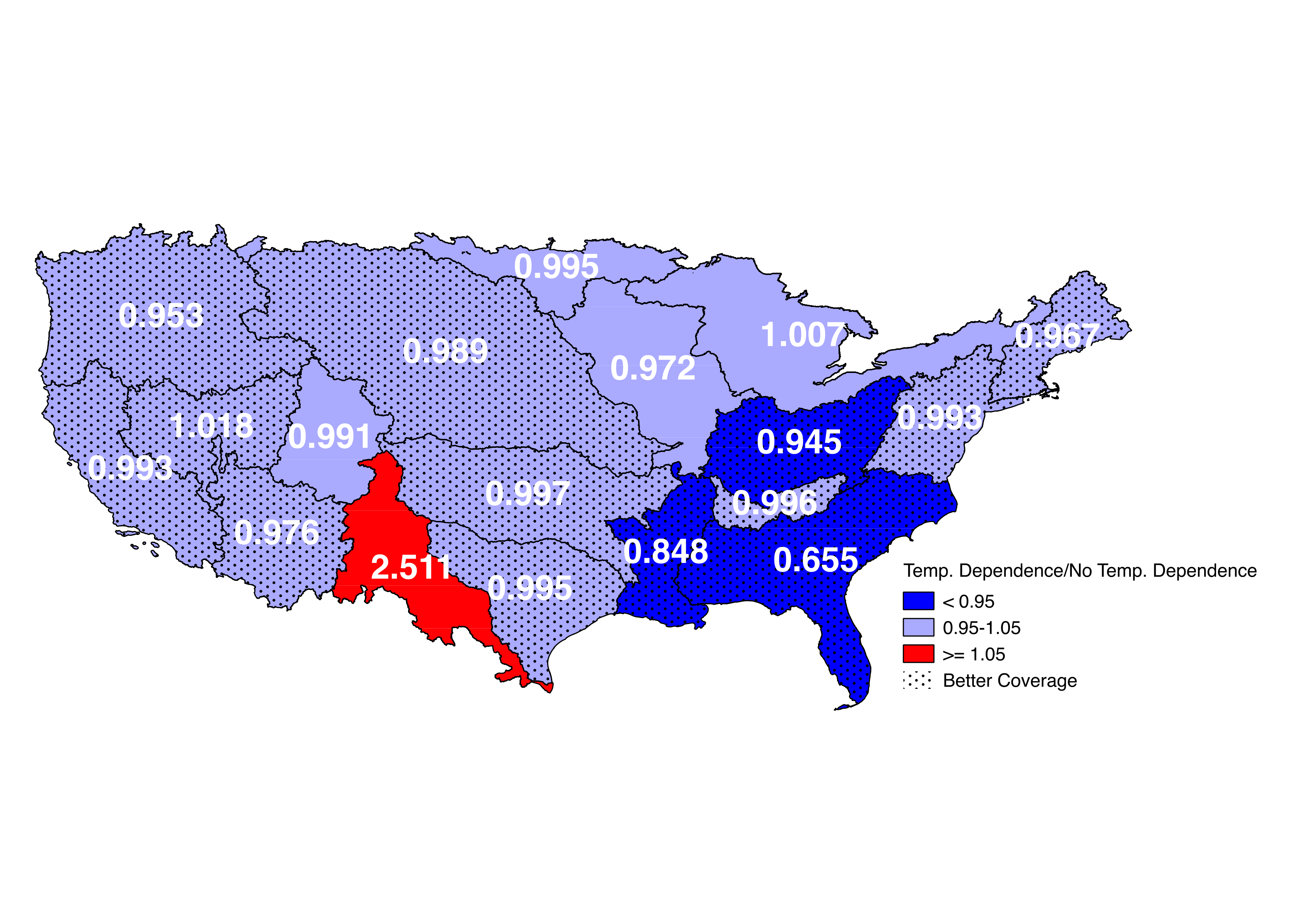}
	\caption{As described further in section \ref{results}\ref{tempcovariate}, the Bayesian model was run once with temperature dependence (with $\phi_{m}$ and $\phi'_{m}$ set as unknown parameters) and once excluding temperature dependence (with $\phi_{m}$ and $\phi'_{m}$ fixed at 0) for each watershed. Watersheds are colored and labeled according to the ratio of  $\frac{RMSE_{p, \phi}}{RMSE_{p, !\phi}}$, where $\phi$ means temperature dependence is included and $!\phi$ means it is not. Watersheds are labeled according to the same ratio. In 15 of 18 watersheds, RMSE is smaller for the model with temperature dependence, though often not substantially different. The most notable exception is the Rio Grande watershed, where the model with temperature results in a significantly larger RMSE. On the other hand, the Lower Mississippi, South Atlantic Gulf, and Ohio watersheds show substantially better RMSE when including temperature covariance. Stippling indicates where coverage when including $\phi$ and $\phi'$ is greater than or equal to coverage without those parameters. In 13 of 18 regions, the model with temperature dependence provides higher coverage. The model with temperature dependence exhibits better upper coverage than the model without it in all 18 cases. }
	\label{phiexperimentmap}
\end{figure}

\begin{figure}[h]
\centering
	\includegraphics[width=\linewidth]{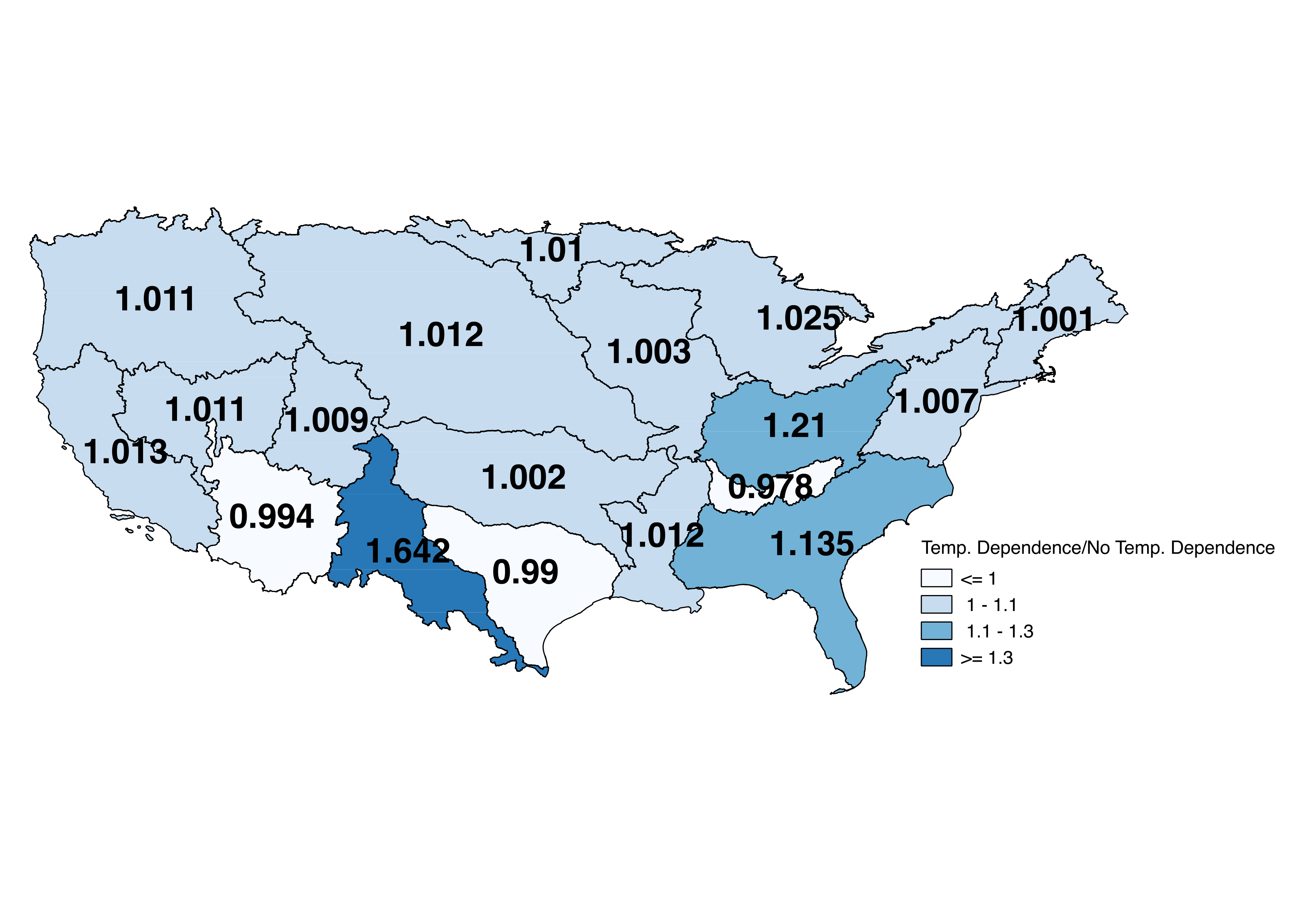}
	\caption{Watersheds are colored and labeled according to the ratio of  $\frac{W_{p, \phi}}{W_{p, !\phi}}$, where $W_{p, \phi}$ means width when temperature dependence is included and $W_{p, !\phi}$ means width when it is not. The blue color palette was chosen for this map as a neutral gradient, since the optimal values of $W_{p}$ depend on the context of the accuracy and posterior coverage attributes as shown in Figure \ref{phiexperimentmap}. Smaller values of $W_{p}$ are only ideal if the Bayesian model also exhibits high coverage, lest the bounds be overly narrow. It may be desirable in cases that $W_{p}$ be larger to reflect larger and potentially irreducible uncertainty. In 16 of 18 cases, the model run with temperature dependence shows a wider posterior distribution.}
	\label{phiexperimentwidthmap}
\end{figure}

Figures \ref{phiexperimentmap} and \ref{phiexperimentwidthmap} synthesize the relative difference in model performance when including temperature dependence versus not including it. Including temperature dependence has a positive or neutral effect on posterior distribution upper bounded coverage in most (16 of 18) cases. In most cases (15 of 18), including temperature dependence improves accuracy in terms of RMSE, though usually not substantially. In the Lower Mississippi, South Atlantic-Gulf, and Ohio watersheds, accuracy improves notably with temperature dependence included. For example, literature has shown that El Ni\~no temperature influences heavy rainfall in the southeastern United States \citep{gershunov1998enso}. In contrast, on the west coast, temperature dependence seems to generally have little effect on results; this could relate to research showing more complex relationships (e.g., the so called Pineapple Express) between oscillations, associated temperature patterns, and heavy precipitation \citep{higgins2000extreme}. 

\section{Discussion} \label{discussion}
In this study, we present a physics-guided Bayesian model that exploits ensembles of ESMs to estimate probabilistic projections of precipitation extremes under climate change. We exploit the knowledge that in many regions there is a relationship between temperature and extreme precipitation (e.g., \cite{pall2007testing}, but that the dependence structure between the two variables might often be more complex than idealistic Clausius Clapeyron scaling \citep{wang2017}. The Bayesian model weights ESMs according to their ability to capture not only historically observed marginal, univariate statistics of daily total precipitation return levels but also their covariance with historically observed same-day average surface temperature. This is an extension of an existing skill- and consensus-based Bayesian ESM weighting framework (e.g., \cite{tebaldi2004regional, tebaldi2005quantifying}). This study has a similar goal to a Bayesian model for precipitation extremes developed recently \citep{sunyer2014bayesian} but is more generalized in the sense that it simultaneously models multiple order statistics of extreme events rather than one specific statistic of extreme precipitation. 

For the model specific to precipitation extremes developed here, there are several caveats worth highlighting. Current generation ESMs do not explicitly resolve phenomena like (extra)tropical cyclones, and thus extreme precipitation as a result of those types of events in the observational record might not be expected to be reflected directly by ESMs. This may impact the the Bayesian model's ability to accurately capture some of the most extreme observed events (e.g., see Figure \ref{quantile_rmse_heatmap}). 

As discussed in section \ref{background}\ref{physics}, there is the potential for nonlinear dependence between temperature and extreme precipitation (Wang et al. 2017). However, with the relatively small number ($Q=Q'=25$) of return levels used here for each unique combination of season and watershed, nonlinear dependence would be difficult to encode into the Bayesian architecture. This could potentially be addressed by modeling a much more complete distribution of precipitation and its extremes, but that exercise would likely be accompanied with statistical challenges related to, for instance, ensuring independence of events and/or the need to explicitly encode and model serial dependence among events. 

The assumptions of serial independence, stationarity, and normality of transformed precipitation return levels are also in select cases questionable, as discussed more in the Appendix. There may be complex relationships between the degree to which climate data meets these assumptions and the effectiveness of, for example, certain prior parameter value choices or Bayesian model design choices in general. As discussed more in the Appendix, certain combinations of prior parameters may work particularly well in certain watersheds, but in this study we opted to find one set of parameters that worked well, generally. It is clear, however, from a prior sensitivity analysis (see Appendix) that results can be quite sensitive to prior choices. Priors that influence results substantially (especially $\iota_0$ and $\kappa_0$ in this study) could theoretically be treated with hyperprior parameters; in practice, anecdotally, we observed that trying this made chain convergence difficult given that those parameter updates were no longer Gibbs sampling steps. Future research may be needed to explore this topic more.

As discussed in section \ref{background}\ref{skillvsconsensus}, the Bayesian model performs best when skill is favored over consensus. This may suggest that in general, weighting ESMs based on nontrivial and physics-guided measures of historical skill (in this case, how well ESMs portray precipitation-temperature dependence from observed data) can lead to improvements in the statistical attributes of probabilistic projections, e.g., accuracy and coverage.

We propose that the model built in this study could perhaps adapted to model any generic set of univariate or multivariate order statistics of outputs from ESMs, whether those statistics are only extremes or more generally order statistics from any climate variable(s) (e.g., the full distributions of temperature, precipitation, humidity, wind speed). In addition, it may be possible to leverage multivariate normal Bayesian data models to share information across space, to quantify correlation among ESMs, and/or to quantify correlation among multiple climate variables. For the purposes of developing stakeholder oriented tools such as complete, climate change scenario-conditioned IDF curves \citep{mailhot2007assessment, mirhosseini2013impact, shrestha2017developing}, it may also be necessary to consider modeling climate variables on sub-daily time scales or on longer aggregated (e.g., 3-day precipitation) ones. Future studies should explore these potential extensions of the model proposed in this study.

%%%%%%%%%%%%%%%%%%%%%%%%%%%%%%%%%%%%%%%%%%%%%%%%%%%%%%%%%%%%%%%%%%%%%
% ACKNOWLEDGMENTS
%%%%%%%%%%%%%%%%%%%%%%%%%%%%%%%%%%%%%%%%%%%%%%%%%%%%%%%%%%%%%%%%%%%%%
%
\acknowledgments
Funding was provided by NSF Expeditions award number 1029711 (Kodra, Chatterjee, Ganguly), NSF SBIR award number 1621576 (Kodra), NSF Big Data award number 1447587 (Ganguly), NSF Cyber SEES award 1442728 (Ganguly), a grant from NASA AMES (Ganguly), and NSF Division Of Mathematical Sciences award number 1622483 (Chatterjee). Kodra is a principal of risQ, Inc., a private for-profit company. Chatterjee and Ganguly are advisers of and shareholders to risQ, Inc. Chen is an intern at risQ, Inc. 

%%%%%%%%%%%%%%%%%%%%%%%%%%%%%%%%%%%%%%%%%%%%%%%%%%%%%%%%%%%%%%%%%%%%%
% APPENDIXES
%%%%%%%%%%%%%%%%%%%%%%%%%%%%%%%%%%%%%%%%%%%%%%%%%%%%%%%%%%%%%%%%%%%%%
%
\appendix
\appendixtitle{Appendix}
\subsection{Data and Preprocessing: Additional Detail} \label{preprocessingdetail}
Table \ref{esms} summarizes the ensemble of 15 ESMs from the CMIP5 archive used in this study. For the years 1950-1999, historical ESM runs are used. For the years 2065-2089, runs from the greenhouse gas emissions scenario Relative Concentration Pathway (RCP) 8.5 are used.
The Bayesian model presented in the main text is run for all 18 continental U.S. Hydrologic Unit 2 (HU2) watersheds provided by the United States Geological Survey's Watershed Boundary Dataset (USGS WBD) \citep{berelson2004mapping}. Table \ref{watersheds} lists and provides a short identification number for each of the 18 watersheds.

\subsection{Priors} \label{priors}
Conjugate prior distributions are defined as follows:

\begin{equation} \label{priorscprimem}
P(C'_m) \sim N(C_{0},\sigma_{0}^{-1}) 
\end{equation}

\begin{equation} \label{priorscbiasj}
P(CBIAS_j) \sim N(\delta_0,\rho_0^{-1})
\end{equation}

\begin{equation} \label{priorssigmaj}
P(\sigma_j) \sim G(\alpha_0, \beta_0)
\end{equation}

\begin{equation} \label{priorstheta}
P(\theta) \sim G(\zeta_0, \eta_0)
\end{equation}

\begin{equation} \label{priorsgammaprimemqprime}
P(\gamma'_{m, q'}) \sim N(\gamma_0,\epsilon_0^{-1})
\end{equation}

\begin{equation} \label{priorsalphajm}
P(\alpha_{j,m}) \sim N(\nu_0,\omega_0^{-1})
\end{equation}

\begin{equation} \label{priorsphi}
P(\phi_{m}, \phi'_{m}) \sim N(\phi_0,\xi_0^{-1})
\end{equation}

\begin{equation} \label{priorsepsilonjq}
P(\epsilon_{j,q}) \sim G(\lambda_0, \upsilon_0)
\end{equation}

\begin{equation} \label{priorsbetaprimemqprime}
P(\beta'_{m, q'}) \sim G(\kappa_0, \iota_0)
\end{equation}

\subsection{Prior Choices and Sensitivity} \label{priorchoices}
Our choices for priors values, with justifications if notable, are shown in Table \ref{priorvalues}. Table \ref{selectedpriorvalues} defines the values we explore for the priors that exert relative influence, those marked with ** in Table \ref{priorvalues}. Note that these are all priors for variance scaling parameters. In Appendix section \ref{mcmc}, we provide more general details on the MCMC procedure. However, owing to computational time considerations, for the above experiment (where there are 625 combinations of the 4 priors parameters in total, for each watershed), we reduce iterations to $N_{1}=300$ for the burnin and $N_{final}=1,000$ after thinning.

\begin{figure}[h]
	\centering
	\includegraphics[width=0.75\textwidth]{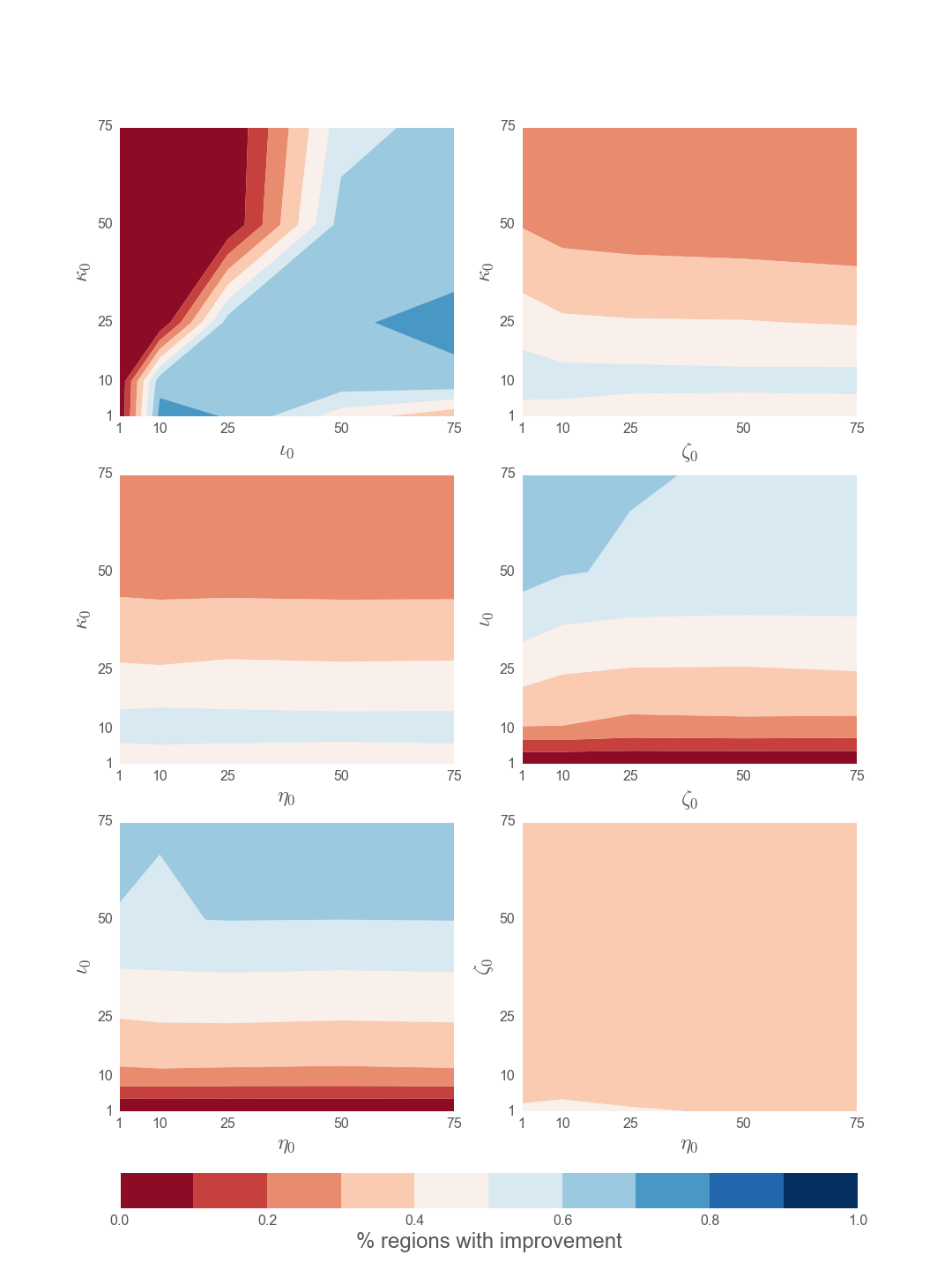}
	\caption{Prior sensitivity is examined across all 18 HU2 watersheds for the parameters $\kappa_0$,  $\iota_0$, $\eta_0$, and $\zeta_0$. Percent of regions where $RMSE_{p} \leq RMSE_{e}$ is depicted with the contour plots for all pairwise combinations of those four parameters. Choice of $\kappa_0$ and $\iota_0$ exert the largest influence over model performance; this can be seen most clearly in the upper left plot.}
	\label{priorexperiment}
\end{figure}

Figure \ref{priorexperiment} shows the results from this sensitivity test. The Bayesian model accuracy is most sensitive to choices of $\kappa_0$ and $\iota_0$, as can be seen most clearly from the top left panel. In all other experiments and in the final model runs, based on the results from this experiment, we set the 4 selected priors as: $\kappa_0=10$, $\iota_0=25$, $\zeta_0=1$, and $\eta_0=10$, for every watershed. These choices can be interpreted in post-hoc fashion to an extent. Setting $\kappa_0 < \iota_0$ means that our prior expectation is that $\beta_{m, q'} < 1$, or that consensus should be favored less than skill in choosing weights $\epsilon_{j,q}$. But the actual absolute values of these priors and the specific ratio of $\frac{\kappa_0}{\iota_0}$ that work well in terms of validation metrics (see Appendix section \ref{validation}) appear to depend on the absolute value range of the precipitation data itself, and as such we our final choices for these two parameters are informed by this experiment. Similarly, $\zeta_0 < \eta_0$ means that our prior belief is that $\theta < 1$, again that consensus should be favored less than skill in choosing weights $\sigma_{j}$.

It is worth noting that different choices of these selected priors, custom-selected per watershed, can lead to improved results in terms of validation. This was observed anecdotally when exploring results from this experiment. We purposefully refrained from choosing different priors per watershed in this study in an effort to not ``overfit" and to avoid losing the value of having one set of interpretable priors. We also note the caveat that this experiment violates the principle of prior parameter selection, which is conventionally not supposed to be informed by data. An alternative approach could have been treating the priors themselves are random parameters with hyperpriors. We experimented informally with this approach; practically speaking, the MCMC chain took a long time to (questionably) approach convergence given that the MCMC update steps for these priors are not Gibbs steps \citep{casella1992explaining}.

\subsection{Posteriors} \label{posteriors}
Full conditional posterior distributions are shown as follows:

% C'_m
\begin{equation} \label{postcprimem}
P(C'_m | ...) \sim N(\frac{c'}{d'}, \frac{1}{d'})
\end{equation}
where
\begin{equation}
c'=\sigma_{0}C_{0}   +  \theta\sum_{j}^{} \sigma_{j}(Z'_{j,m} - CBIAS_{j}) 
\end{equation}
and
\begin{equation}
d'=\sigma_{0}   +  \theta\sum_{j}^{} \sigma_j
\end{equation}

% CBIAS_j
\begin{equation} \label{postcbiasj}
P(CBIAS_j | ...) \sim N(\frac{g}{h}, \frac{1}{h})
\end{equation}
where
\begin{equation} 
g=\rho_0\delta_0 +  \sigma_{j}\sum_{m}^{}(Z_{j,m}- C_m) + \theta\sigma_{j}\sum_{m}^{}(Z'_{j,m} - C'_m)
\end{equation}
and
\begin{equation} 
h=\rho_0 +  M\sigma_j(1 + \theta) 
\end{equation}

% sigma_j 
\begin{equation} \label{postsigmaj}
P(\sigma_j |...) \sim G(g, h)
\end{equation}
where
\begin{equation}
g=\alpha_0 + M
\end{equation}
and
\begin{equation}
\begin{split}
h = \beta_0 + 0.5\sum_{m}^{}((U_{j,m,q=1} - C_{m} - CBIAS_{j})^{2}) + \\
   0.5\theta\sum_{m}^{}((U'_{j,m,q'=1} - C'_{m} - CBIAS_{j})^{2}) 
\end{split}
\end{equation}

% theta
\begin{equation} \label{posttheta}
P(\theta |...) \sim G(n, o)
\end{equation}
where
\begin{equation}
n=\zeta_0 + 0.5J + 0.5M
\end{equation}
and
\begin{equation} 
o = \eta_0 +  0.5\sum_{j,m}^{}(\sigma_j(U'_{j,m,q'=1} - C'_{m} - CBIAS_{j})^{2})
\end{equation}

The next posteriors for $q >1$ and $q'>1$ can be estimated completely independently of those above for $q=q'=1$, which can be advantageous computationally:

% gamma'_{m,q'}
\begin{equation} \label{postgammaprimemqprime}
P(\gamma'_{m,q'} |...) \sim N(\frac{p'}{r'}, \frac{1}{r'}) 
\end{equation}
where
\begin{equation}
\begin{split}
p'=\epsilon_0\gamma_0 + \beta'_{m,q'}\sum_{j}^{}\epsilon_{j,q}(U'_{j,m,q'} - \alpha_{j,m} - \phi'_m\delta'_{j,m,q'})
\end{split}
\end{equation}
and
\begin{equation}
r=\epsilon_0 + \beta'_{m,q'}\sum_{j}^{}\epsilon_{j,q}
\end{equation}

%phi_m 
\begin{equation} \label{postphim}
P(\phi_{m} | ...) \sim N(cc, dd)
\end{equation}
where
\begin{equation}
\begin{split}
cc=\xi_0\phi_0 + \sum_{k,q}^{}\tau_{k}\epsilon_{k,q}(\delta_{k,m,q})(U_{k,m,q} - \gamma_{m,q}) + \\
      \sum_{j,q}^{}\epsilon_{j,q}(\delta_{j,m,q})(U_{j,m,q} - \gamma_{m,q} - \alpha_{j,m})
\end{split}
\end{equation}
and

\begin{equation}
\begin{split}
dd=\xi_0 +  \sum_{k,q}^{}\tau_{k}\epsilon_{k,q}(\delta_{k,m,q})^2 + \sum_{j,q}^{}\epsilon_{j,q}(\delta_{j,m,q})^2
\end{split}
\end{equation}

%phi'_m
\begin{equation} \label{postphiprimem}
P(\phi'_{m} | ...) \sim N(cc', dd')
\end{equation}
where
\begin{equation} 
cc=\xi_0\phi_0 + \sum_{j,q'}^{}\beta'_{m,q'}\epsilon_{j,q}(\delta_{j,m,q'})(U'_{j,m,q'} - \gamma'_{m,q'} - \alpha_{j,m})
\end{equation}
and
\begin{equation} 
\begin{split}
dd=\xi_0  + \sum_{j,q'}^{}\beta'_{m,q'}\epsilon_{j,q}(\delta'_{j,m,q'})^2
\end{split}
\end{equation}

%epsilon_j_q
\begin{equation} \label{postepsilonjq}
 P(\epsilon_{j,q}|...) \sim G(ii, jj)
 \end{equation} 
 where
 \begin{equation} 
ii = \lambda_0 + M
\end{equation} 
and 
\begin{equation} 
\begin{split}
jj = \upsilon_0 + 0.5\sum_{m}^{}(U_{j,m,q} - \gamma_{m,q} - \alpha_{j,m} -  \phi_m\delta_{j,m,q})^{2} + \\
 0.5\sum_{m}^{}\beta'_{m,q'}(U'_{j,m,q'} - \gamma'_{m,q'} -  \alpha_{j,m} -  \phi'_m\delta'_{j,m,q'})^{2}
\end{split}
\end{equation} 

%beta_m,q'
\begin{equation} \label{postbetaprimemqprime}
P(\beta'_{m,q'}|...) \sim G(rr', ss')
\end{equation}
where
\begin{equation}
rr= \kappa_0 + 0.5J
\end{equation}
and
\begin{equation} 
ss = \iota_0 + 0.5\sum_{j}^{}\epsilon_{j,q}(U'_{j,m,q'} - \gamma'_{m,q'} - \alpha_{j,m} -  \phi'_m\delta'_{j,m,q'})^{2}
\end{equation}

% alpha_j,m
\begin{equation} \label{postalphajm}
P(\alpha_{j,m} |...) \sim N(u, v)
\end{equation}
where
\begin{equation}
\begin{split}
u=\omega_0\nu_0 + \sum_{q}^{}\epsilon_{j,q}(U_{j,m,q} - \gamma_{m,q} - \phi_m\delta_{j,m,q})  \\
  + \sum_{q'}^{}\beta'_{m,q'}\epsilon_{j,q}(U'_{j,m,q'} - \gamma'_{m,q'} - \phi'_m\delta'_{j,m,q'})
\end{split}
\end{equation}
and
\begin{equation}
v=\omega_0 + \sum_{q}^{}\epsilon_{j,q} + \sum_{q'}^{}\beta'_{m,q'}\epsilon_{j,q}
\end{equation}

\subsection{Markov Chain Monte Carlo and Diagnostics}\label{mcmc}
Since all priors are conjugates, all posteriors of the Bayesian model can be estimated through a Gibbs sampler \citep{casella1992explaining}, where each unknown is iteratively sampled conditional on the current values of all other unknown parameters. 

In each simulation, all unknowns must be initialized. In theory, the MCMC chain should converge to the true target joint distribution regardless of the initial values of the unknowns. However, in order to encourage fast practical convergence, we aim to select well-reasoned values. Table \ref{startingvaluestable} provides the selected starting values for each unknown parameter.

Some values may be relatively far away from the center of their true distributions. Following previous related literature (Tebaldi et al. 2005), we initially ran MCMC runs with burnins of size $N_{1}=250,000$ followed by $N_{2}=250,000$ more samples. We thinned the chain of size $N_2=250,000$ by only saving every 50th posterior sample to induce independence between samples, also following the same literature (Tebaldi et al. 2005). This provided a final posterior of size $N_{final}=5,000$. 

We compared results from this setup to a less computationally expensive one. Specifically, the final default setup for MCMC runs was a burnin of size $N_1=500$, a second chain of variable size $N_2$, and $N_{final}=10,000$. We achieved $N_{final}=10,000$ by thinning the $N_2$ size chain proportional to the effective sample size $N_{eff}$ \citep{sturtz2005r2winbugs} of the burnin sample of $\gamma_{m, q'}$, averaged over all $m$ and $q'$. Specifically, we rounded the ratio $\frac{N_1}{N_{eff}}$ to the nearest integer and took the minimum of this ratio or 10 as a thinning constant $TC$. Then, to generate $N_{final}=10,000$, we saved every $TC^{th}$ sample from the post-burnin chain. In practice, we found no notable difference in posterior results between this setup and the one from previous literature (Tebaldi et al. 2005), hence we used this as our default setup unless otherwise noted.  

We run standard MCMC diagnostics to check independence of samples and approximate chain convergence. Using the same procedure as above, we again compute the effective sample size of this final thinned posterior sample $N_{final}$. Those effective sample sizes for the validation scheme runs are shown in Table \ref{mcmcdiagnosticstable}; most are very close to 10,000.

We also use the Geweke diagnostic \citep{geweke_evaluating_1991} to assess whether the chain has converged to the target posterior distribution. The Geweke diagnostic assumes that the last half of the chain has converged to the target distribution. If the mean of an earlier portion (here, the first $5,000$ values of the final chain) is not significantly different from the mean of the last half (the last 5,000), then it can be reasonable inferred that the chain converged in that portion or earlier \citep{geweke_evaluating_1991}. The Geweke diagnostic is computed for each component $m$, $q'$ of the final posterior distribution of the unknown $\gamma'_{m,q'}$. The diagnostic is a test statistic is a standard Z-score that represents the difference between the two sample means divided by its estimated standard error. The standard error is estimated from the spectral density at zero and so takes into account any residual autocorrelation after thinning (Geweke 1991). Since we do this over all $m$ and $q'$ (in our case a total of $12 x 25 = 300$ times), we could expect for example $5\%$ of Z-scores to exceed $1.96$ in absolute value by chance. Further, since each value of $\gamma'_{m,q'}$ is updated using shared information across $M$ and $Q'$, it could be reasonable to see correlation of Z-scores over seasons and over ordered return levels. This dependence between tests could distort that $5\%$ percent expectation. For each watershed, we tabulate the percentage of tests where Geweke Z-scores exceed 1.96 in absolute value in Table \ref{mcmcdiagnosticstable}.

\subsection{Statistical assumptions} \label{assumptions}
The design of the Bayesian model involves several statistical assumptions that we define here. 

\begin{figure}[h]
\centering
	\includegraphics[width=\linewidth]{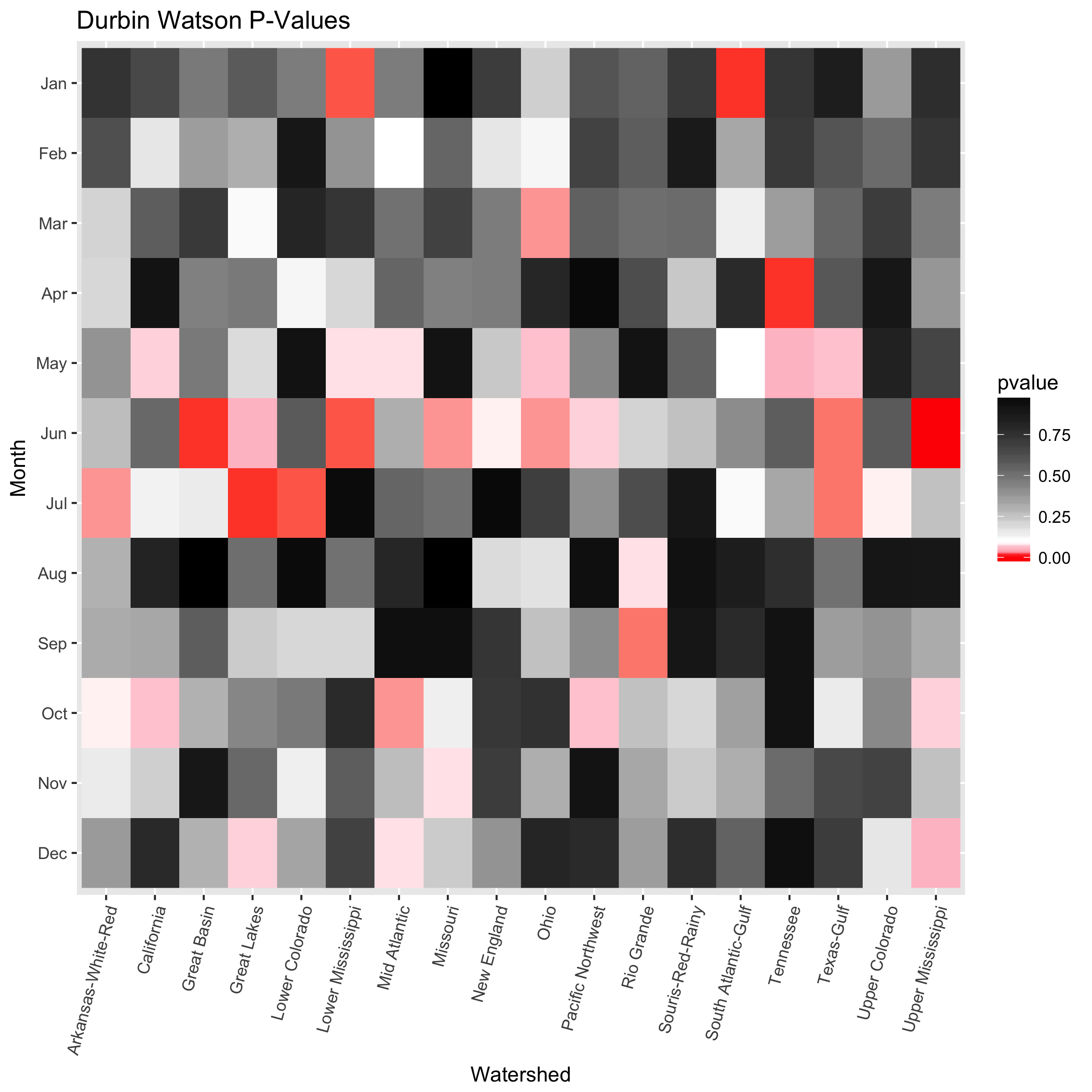}
	\caption{Prior to reordering precipitation return levels in ascending order of intensity, the Durbin-Watson serial dependence test is applied in each watershed for observational data return levels with respect to each season $m$ for the climatology 1950-1974. A heatmap of the Durbin-Watson test statistic p-values is displayed by watershed and month. Red coloring indicates significance at 0.05, pink at 0.10, and grayscale is used for p-values above 0.10. }
	\label{DWPvalue_heatmap}
\end{figure}

\textbf{Serial Independence} - First, by re-ordering the original block maxima (return levels) in ascending order of intensity, we are effectively making the assumption that there is no serial correlation between temporally ordered return levels. This is a typical assumption in extreme value modeling of block maxima (Coles et al. 1991). The assumption allows us to avoid modeling temporal dependence. To check this assumption, prior to reordering or data processing, we employ the Durbin-Watson serial dependence test \citep{durbin_testing_1952} in each watershed for return levels of observational data with respect to each season $m$ for the climatology 1950-1974. Figure \ref{DWPvalue_heatmap} displays a heatmap of the Durbin-Watson test statistic p-values for each watershed and season $m$. In most cases, p-values are not significant at a 0.05 level. However, the distribution is not quite uniform; 19 ($\sim9\%$) p-values are $<= 0.05$, where we would only expect 5$\%$ of p-values to be significant by chance. In particular, statistically significant serial correlations load heavily onto the months of June and July. 

\begin{figure}[h]
\centering
	\includegraphics[width=\linewidth]{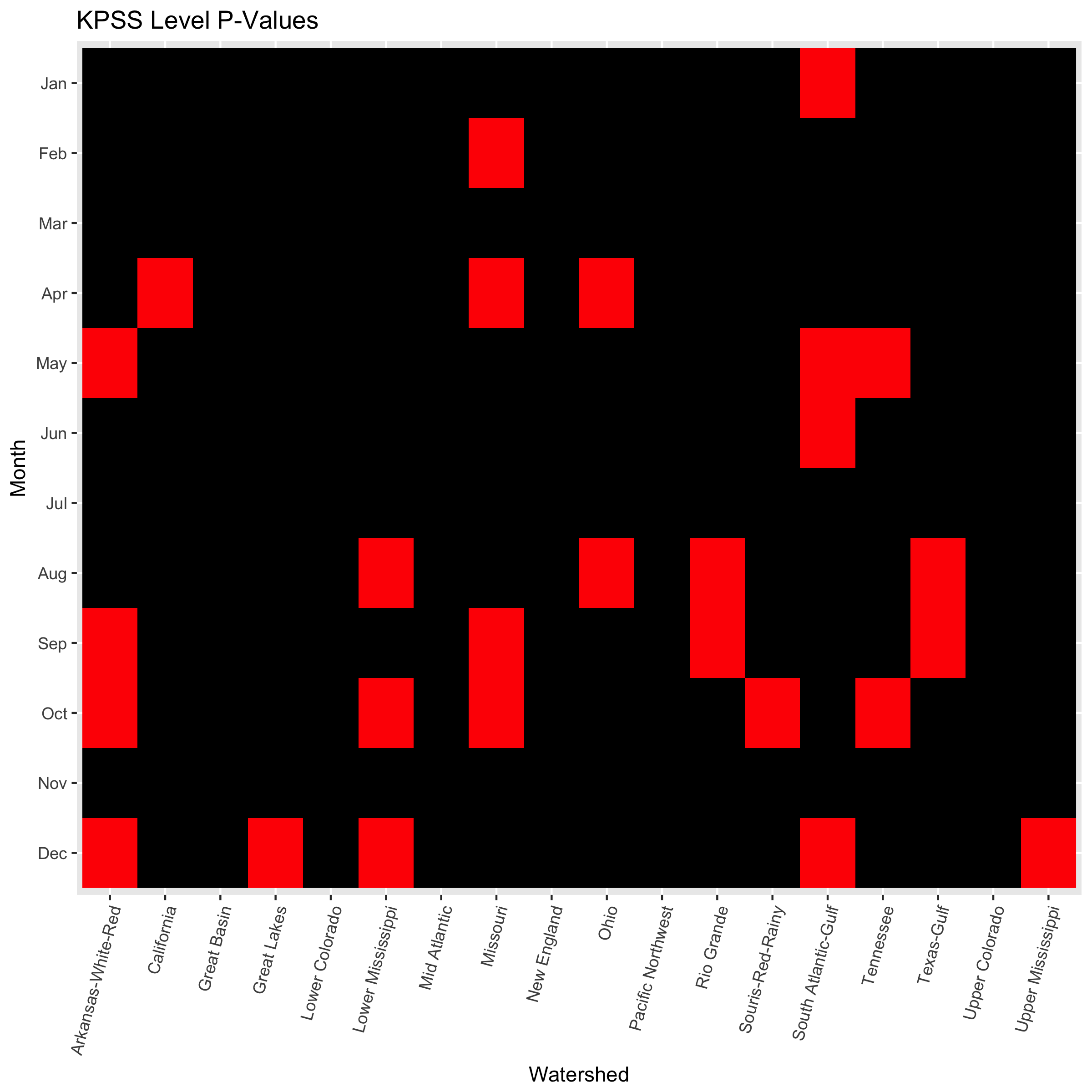}
	\caption{Prior to reordering precipitation return levels in ascending order of intensity, KPSS tests are employed to test for level stationarity. The tests are applied in each watershed for block maxima observational data with respect to each season $m$ for the climatology 1950-1974. In this heatmap, red coloring indicates significant level non-stationarity (i.e., the mean of the time series is not constant) at a 0.05 level. Black indicates p-values that are not significant at 0.05. }
	\label{KPSSLevelPvalue_heatmap}
\end{figure}

\begin{figure}[h]
\centering
	\includegraphics[width=\linewidth]{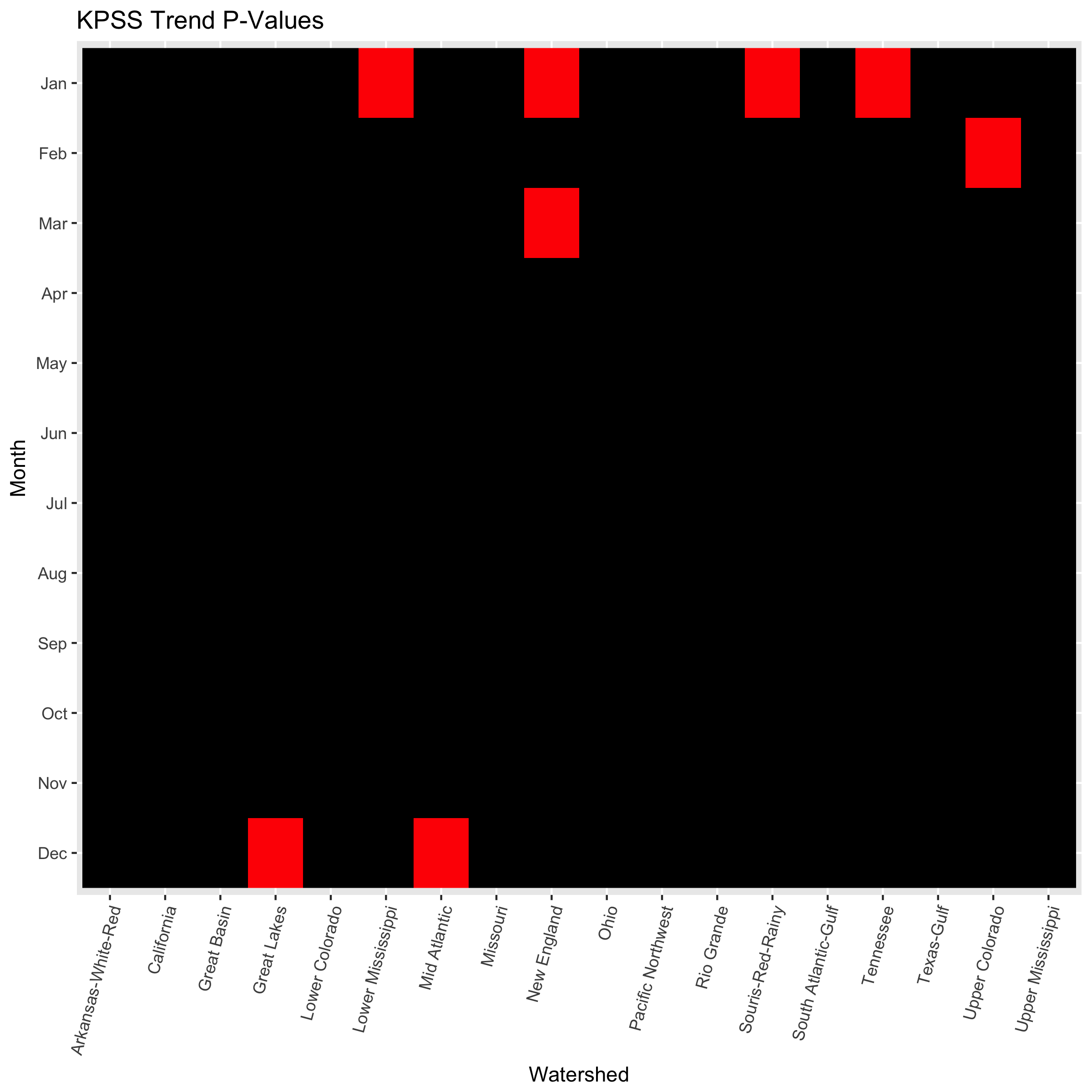}
	\caption{The same as in Figure \ref{KPSSLevelPvalue_heatmap} is shown but for KPSS trend stationarity tests.}
	\label{KPSSTrendPvalue_heatmap}
\end{figure}

\textbf{Stationarity} - As within a standard extreme value modeling setting, time-ordered return levels are assumed to be level and trend stationary \citep{kwiatkowski_testing_1992}. We utilize the Kwiatkowski - Phillips - Schmidt - Shin (KPSS) tests for both types of stationarity in each watershed for return level observations with respect to each season $m$. Figures \ref{KPSSLevelPvalue_heatmap} and \ref{KPSSTrendPvalue_heatmap} displays a heatmap of the KPSS level and trend stationarity test outcomes, respectively, similar to Figure \ref{DWPvalue_heatmap} for the Durbin-Watson tests. R's base library KPSS test function reports any p-values $>= 0.10$ simply as being $>= 0.10$, so these heatmaps only delineate the significant versus insignificant test statistics at 0.05. For the level stationarity test, 27 (12.5$\%$) of cases are significant, more than would be expected by chance. Meanwhile only 8 ($\sim3.7\%$) of KPSS trend stationarity tests are significant. Overall these test results may not be surprising given that research on extreme precipitation trends in the 20th century \citep{kunkel2003north}.

\begin{figure}[h]
\centering
	\includegraphics[width=\linewidth]{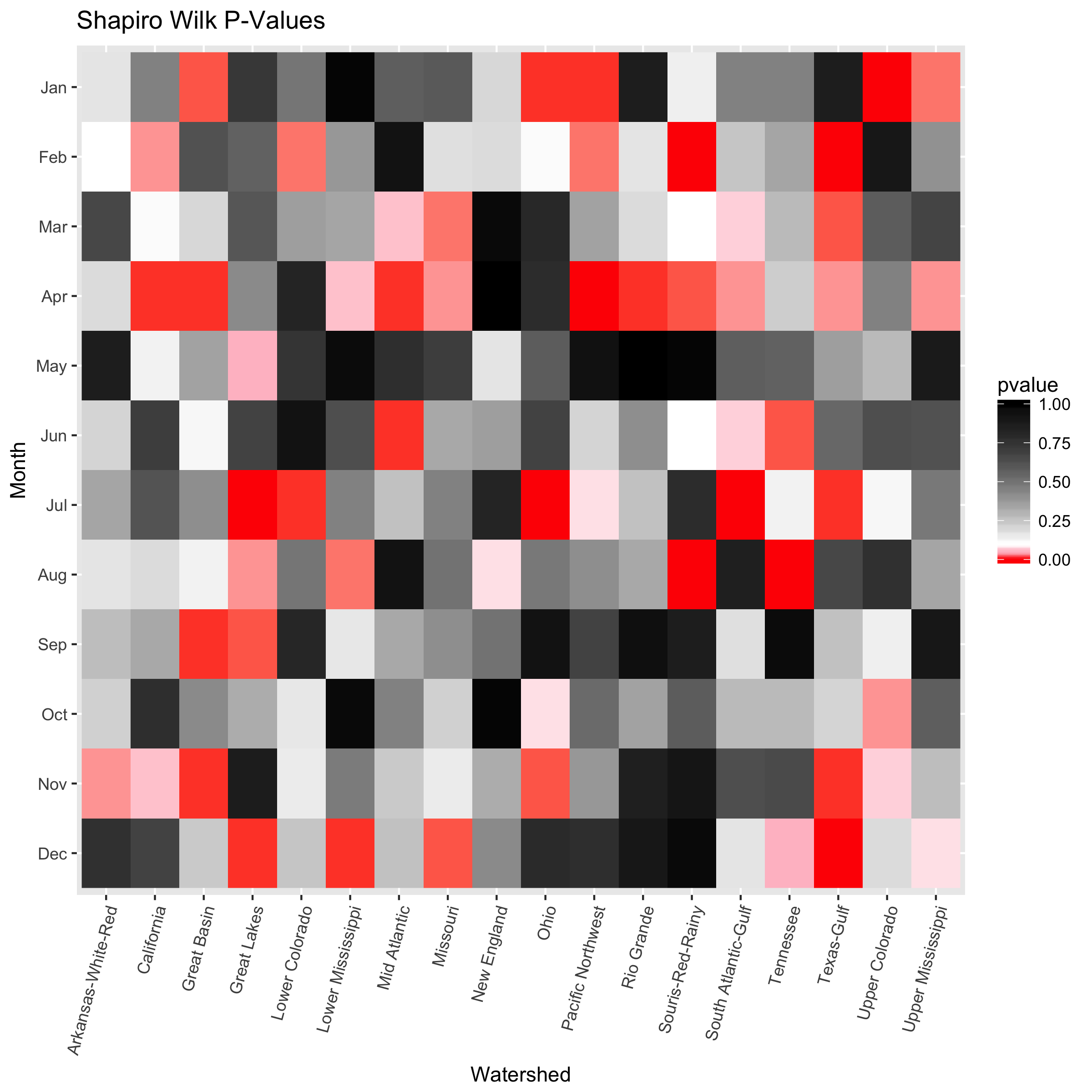}
	\caption{The same is shown as in Figure \ref{DWPvalue_heatmap} but for Shapiro Wilk normality tests.}
	\label{SWPvalue_heatmap}
\end{figure}

\textbf{Normality} - $U_{j,m,q}$ and $U'_{j, m, q'}$ for any ESM $j$ (or observational dataset indexed by $k$) are are assumed to be Gaussian conditional on temperature dependence. Recall $U_{j,m,q}=\log(P_{j,m,q} - P_{j,m,q-1})$ for all $q \in [2, ... Q]$. For each observational dataset over the 1950-1974 climatology ($U_{k,m,q}$) in each watershed, we first utilize the Shapiro Wilk test for normality \citep{shapiro1965analysis}. More specifically, for each season $m$, we fit a ordinary least squares linear regression $U_{k,m,q} \sim \delta_{k,m,q}$. We run the Shapiro Wilk test on the residuals from that regression to approximate testing the normality assumption in the Bayesian model. Figure \ref{SWPvalue_heatmap} shows a heatmap like Figure \ref{DWPvalue_heatmap} but for the Shapiro Wilk tests. In most cases, p-values are not significant at a 0.05 level. However, the distribution is not uniform; 46 ($\sim21\%$) p-values are $<= 0.05$, suggesting there is non-normality in some cases, as we would only expect 5$\%$ of p-values to be significance by statistical chance.

\begin{figure}[h]
\centering
	\includegraphics[width=\linewidth]{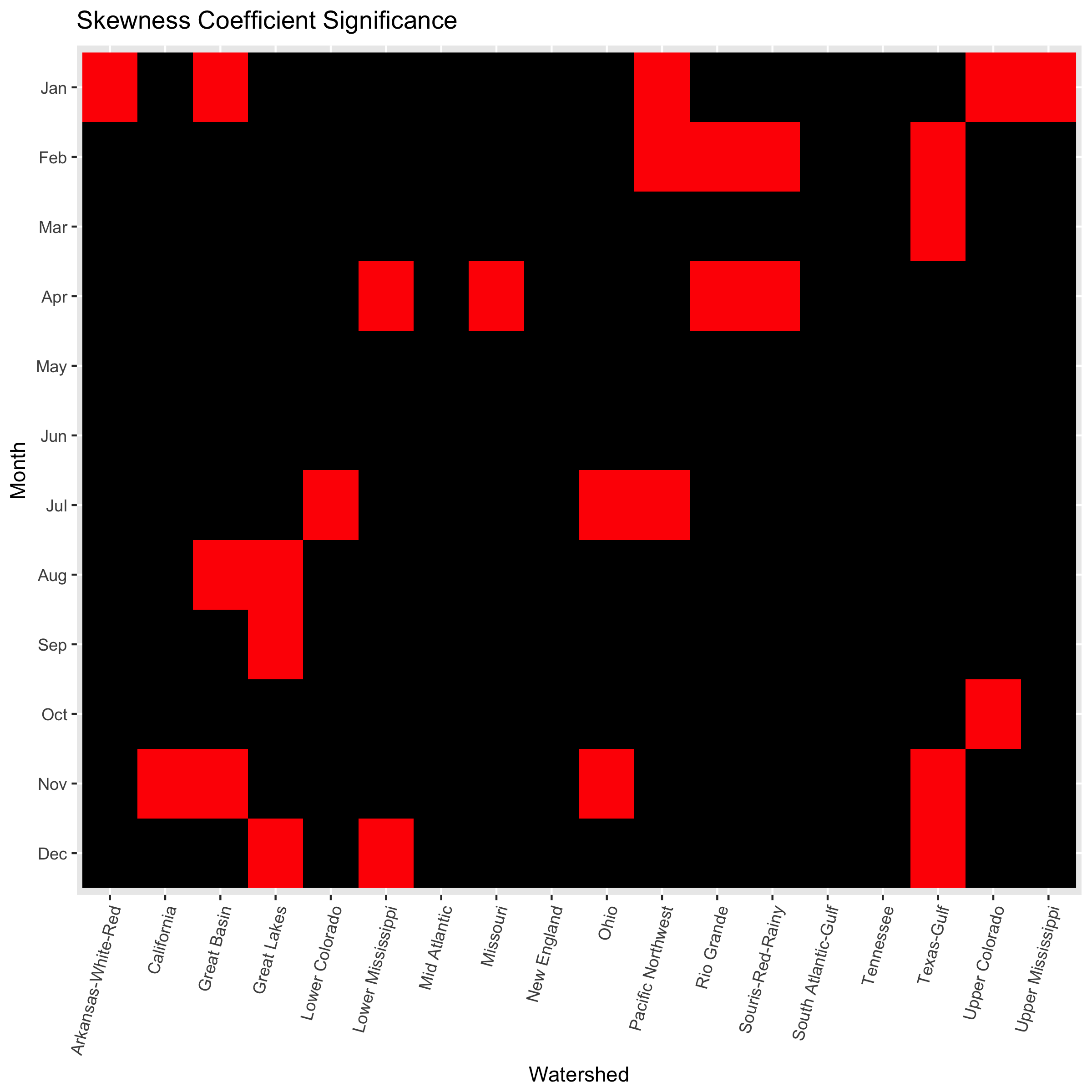}
	\caption{For each month and watershed, $U_{k,m,q}$ for $q \in [2,... Q]$ (observational data from 1950-1974) is utilized to check for significant positive or negative skewness. Specifically, a standard 10,000 member bootstrapped distribution of the sample skewness coefficient is computed from $U_{k,m,q}$. From this a $95\%$ confidence interval is estimated. In every instance where that confidence interval does not include 0, a non-black square is entered in the heatmap. All red squares indicate significant negative skew (where the entire $95\%$ confidence interval is below 0). There are no cases where significant positive skew is found. In total 28 ($\sim 10\%$) of 300 cases show significant negative skewness.}
	\label{skew_heatmap}
\end{figure}

\begin{figure}[h]
\centering
	\includegraphics[width=\linewidth]{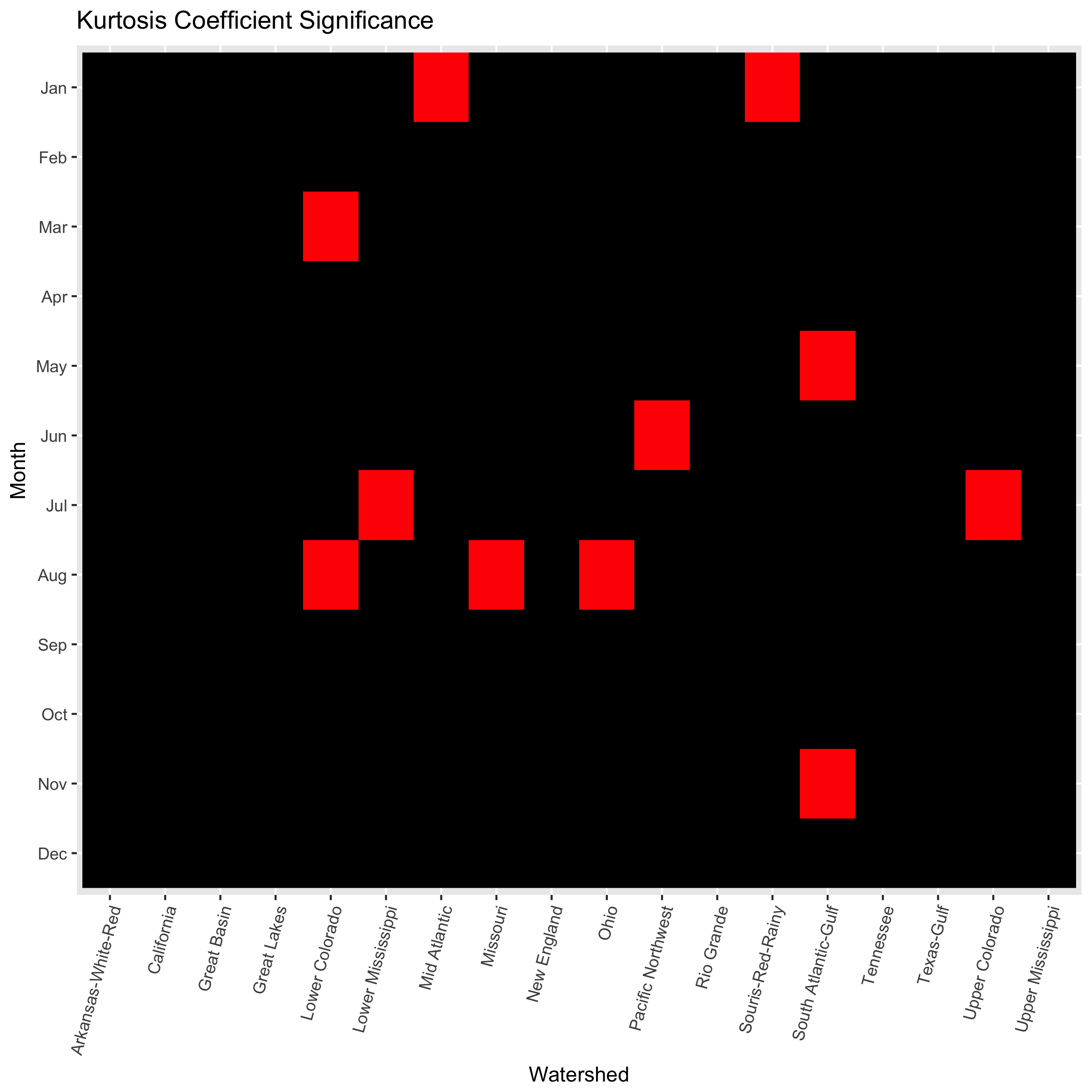}
	\caption{The same is shown as in Figure \ref{skew_heatmap} but for kurtosis. The same bootstrapped values of $U_{k,m,q}$ for $q \in [2,... Q]$ are used for testing significance of both skewness and kurtosis. Only 11 ($\sim 5\%$) of 300 cases show significant negative kurtosis; none show significant positive kurtosis.}
	\label{kurt_heatmap}
\end{figure}

We also complement that test with significance tests for skewness and kurtosis computed on those same residuals \citep{joanes1998comparing}. More specifically, we compute a 95$\%$ confidence interval from a 1000-iteration ordinary bootstrap for sample skewness and kurtosis statistics. We tabulate the occasions when skewness (kurtosis) is significantly negative (positive) based on this bootstrap procedure. In all cases where there is significance in the skewness tests, the statistics are negative, meaning that the distribution of the residuals of $U_{k,m,q}$ after regression on $\delta_{k,m,q}$ is skewed left. This is owing to the log transformation applied to differentials in ordered block maxima, i.e., $\log(P_{k,m,q} - P_{k,m,q-1})$, which sometimes amplifies outlier behavior of small precipitation values, effectively making the left tail more severe. A total of 28 ($\sim13\%$) of cases show significant left skew. Only 11  ($\sim5\%$) of cases exhibit significant negative kurtosis, which falls within the realm of statistical chance. Figures \ref{skew_heatmap} and \ref{kurt_heatmap} show heatmaps for significance in the skewness and kurtosis test statistics, respectively, similar to Figures \ref{KPSSLevelPvalue_heatmap} and \ref{KPSSTrendPvalue_heatmap} for the KPSS test statistics. 

Caution is needed in interpretation in all test results, however, since the tests are not necessarily independent of each other, and behavior of return values could be correlated across watersheds and months. If anything, p-values for all the test results are overly small and, as such, conservative.

%%%%%%%%%%%%%%%%%%%%%%%%%%%%%%%%%%%%%%%%%%%%%%%%%%%%%%%%%%%%%%%%%%%%%
% REFERENCES
%%%%%%%%%%%%%%%%%%%%%%%%%%%%%%%%%%%%%%%%%%%%%%%%%%%%%%%%%%%%%%%%%%%%%
% Make your BibTeX bibliography by using these commands:
\bibliographystyle{ametsoc2014}
\bibliography{kodra_etal_2017}

%%%%%%%%%%%%%%%%%%%%%%%%%%%%%%%%%%%%%%%%%%%%%%%%%%%%%%%%%%%%%%%%%%%%%
% TABLES
%%%%%%%%%%%%%%%%%%%%%%%%%%%%%%%%%%%%%%%%%%%%%%%%%%%%%%%%%%%%%%%%%%%%%
\begin{table}[h]
  \centering
  \caption{Modeling Schemes}
    \begin{tabular}{ | l | l | l | p{5cm} |}
    \hline
    \textbf{Scheme} & \textbf{Historical} & \textbf{Future} \\ \hline
    Validation & 1950-1974 & 1975-1999 \\ \hline
    End Of Century & 1975-1999 & 2065-2089 \\ \hline
    \end{tabular}
     \label{schemes}
\end{table}

\begin{table}[h]
  \centering  
    \begin{tabular}{ | l | l | l |}
    \hline
    \textbf{ESM Full Name} & \textbf{ESM Short Name} & \textbf{Resolution (lon x lat)} \\ \hline
    CanESM2 & canesm2 & 128 x 64 \\ \hline
    CCSM4 & ccsm4 & 288 x 192 \\ \hline
    CESM1-CAM5 & cesm1cam5 & 288 x 192 \\ \hline
    GFDL-CM3 & gfdlcm3 & 144 x 90 \\ \hline
    GFDL-ESM2G & gfdlesm2g & 144 x 90 \\ \hline
    inmcm4 & inmcm4 & 180 x 120 \\ \hline
    IPSL-CM5A-MR & ipslcm5amr & 144 x 143  \\ \hline
    IPSL-CM5B-LR & ipslcm5blr & 96 x 96\\ \hline
    MIROC5 & miroc5 & 256 x 128 \\ \hline
    MIROC-ESM & mirocesm & 128 x 64 \\ \hline
    MIROC-ESM-CHEM & mirocesmchem & 128 x 64 \\ \hline
    MPI-ESM-LR & mpiesmlr & 192 x 96 \\ \hline
    MPI-ESM-MR & mpiesmmr & 192 x 96 \\ \hline
    MRI-CGCM3 &  mricgcm3 & 320 x 160 \\ \hline
    NorESM1-M &  noresm1m & 144 x 96 \\ \hline
    \end{tabular}
    \caption{CMIP5 Earth System Models (ESMs) included in this study. For each ESM, we use only model runs labeled r1i1p1. For all future climatologies, we use ESM outputs conditioned on greenhouse gas trajectory scenario RCP8.5.}
     \label{esms}
\end{table}

\begin{table}[h]
  \centering
  \caption{USGS HU2 Watersheds}
    \begin{tabular}{ | l | l | }
    \hline
    \textbf{ID} & \textbf{Watershed} \\ \hline
    1 & Lower Mississippi \\ \hline
    2 & Tennessee  \\ \hline
    3 & Pacific Northwest  \\ \hline
    4 & Missouri \\ \hline
    5 & Arkansas-White-Red \\ \hline
    6 & Souris-Red-Rainy \\ \hline
    7 & Mid Atlantic \\ \hline
    8 & Upper Colorado \\ \hline
    9 & Lower Colorado \\ \hline
    10 & Ohio \\ \hline
    11 & Upper Mississippi \\ \hline
    12 & New England \\ \hline
    13 & Great Basin \\ \hline
    14 & South Atlantic-Gulf \\ \hline
    15 & Texas-Gulf \\ \hline
    16 & Rio Grande\\ \hline
    17 & California \\ \hline
    18 & Great Lakes \\ \hline
    \end{tabular}
     \label{watersheds}
\end{table}

\begin{table}[h]
\centering
    \begin{tabular}{ | l | l | l | p{5cm} |}
    \hline
    \textbf{Parameter} & \textbf{Value}  & \textbf{Notes} \\ \hline
    $C_{0,m}$ & 0 & \\ \hline
    $\sigma_{0,m}$ &  0.01 & Relatively small weight makes $C_{0,m}$ less informative \\ \hline
    $\delta_{0}$ & 0 & Presume ESMs are unbiased \\ \hline
    $\rho_{0}$ & 1 & \\ \hline
    $\alpha_{0}$ & 0.1 & Uninformative prior for ESM weights   \\ \hline
    $\beta_{0}$ & 0.1 &  \\ \hline
    $\zeta_{0}$ & 1 & ** \\ \hline
    $\eta_{0}$ & 10 & ** \\ \hline
    $\gamma_{0}$ & 1e-7 & Presume a small increase in subsequent block maxima  \\ \hline
    $\epsilon_{0}$ & 1 & \\ \hline
    $\nu_{0}$ & 0 & Presume ESMs are unbiased \\ \hline
    $\omega_{0}$ & 1 & Relatively large weight to influence bias closer to 0 \\ \hline
    $\phi_{0}$ & 0 & Presume no linear relationship with temperature \\ \hline
    $\xi_{0}$ & 1 & Relatively small weight makes $\phi_{0}$ less informative  \\ \hline
    $\lambda_{0}$ & 0.1 & Uninformative prior for ESM weights \\ \hline
    $\upsilon_{0}$ & 0.1 &  \\ \hline
    $\kappa_{0}$ & 10 & ** \\ \hline
    $\iota_{0}$ & 25 & ** \\ \hline    
    \end{tabular}
     \caption{Prior Parameters and Values. Selected priors generally work well in practice. Those rows with ** indicates that we observed relative sensitivity of results to choice of their values. For these, we explore a range of values in a prior sensitivity study in the two regions where we do extensive analysis. Priors without ** exert relatively little influence as long as the choices for their values are reasonable. }     
    \label{priorvalues}
\end{table}

\begin{table}[h]
\centering
  \caption{Selected priors and candidate values.}
    \begin{tabular}{ | l | l | l | p{5cm} |}
    \hline
    \textbf{Parameter} & \textbf{Candidate Values} \\ \hline
    $\zeta_{0}$ & 1, 10, 25, 50, 75  \\ \hline
    $\eta_{0}$ & 1, 10, 25, 50, 75  \\ \hline
    $\kappa_{0}$ & 1, 10, 25, 50, 75 \\ \hline
    $\iota_{0}$ & 1, 10, 25, 50, 75 \\ \hline    
    \end{tabular}
\label{selectedpriorvalues}    
\end{table}

\begin{table}[h]
\centering
  \caption{Parameter Starting Values}
    \begin{tabular}{| l | l | p{8cm} |}
    \hline
    \textbf{Unknown} & \textbf{Value} & \textbf{Notes} \\ \hline
    $C'_m$ & 2 & Starting point of $\exp(2) = \sim 7.4 \frac{mm}{day}$ for smallest block maxima ($q=q'=1$) \\ \hline
    $CBIAS$ & 0 & Presume no ESM bias \\ \hline
    $\sigma_{j}$ & 1 & \\ \hline
    $\theta_{j}$ & 1 & Presume consensus could be equally as important as skill \\ \hline
    $\gamma_{m,q}, \gamma'_{m,q'}$ & 0.05 &\\ \hline
    $\phi_{m}, \phi'_{m}$ & 0 & Presume no temperature-precipitation dependence \\ \hline
    $\alpha_{j,m}$ & 0 & Presume no ESM bias \\ \hline
    $\epsilon_{j,q}$ & 1 &  \\ \hline
    $\beta'_{j,q'}$ & 1 & Presume consensus could be equally as important as skill \\ \hline
    \end{tabular}
    \label{startingvaluestable}
\end{table}

\begin{table}[h]
  \centering
    \begin{tabular}{|l|l|l|l|}
    \hline
    \textbf{ID} & \textbf{Watershed} & \textbf{Effective $N_{final}$} & \textbf{$\%$ Geweke $|Z| \leq 1.96$} \\ \hline
    1 & Lower Mississippi & 9,411 & 100 \\ \hline
    2 & Tennessee & 9,592 & 100 \\ \hline
    3 & Pacific Northwest & 10,000 & 91.3 \\ \hline
    4 & Missouri & 10,000 & 99.3 \\ \hline
    5 & Arkansas-White-Red & 10,000 & 92.3 \\ \hline
    6 & Souris-Red-Rainy & 9,658 & 95 \\ \hline
    7 & Mid Atlantic & 10,000 &  88.7 \\ \hline
    8 & Upper Colorado & 10,000 &  99 \\ \hline
    9 & Lower Colorado & 10,000 &  100 \\ \hline
    10 & Ohio & 9,085 &  91.3 \\ \hline
    11 & Upper Mississippi & 10,000 &  96.3 \\ \hline
    12 & New England & 10,000 &  94.3 \\ \hline
    13 & Great Basin & 9,157 &  99 \\ \hline
    14 & South Atlantic-Gulf & 10,000 & 100 \\ \hline
    15 & Texas-Gulf & 10,000 &  100 \\ \hline
    16 & Rio Grande & 9,671 &  90 \\ \hline
    17 & California & 10,000 &  97.3 \\ \hline
    18 & Great Lakes & 10,000 & 90.7 \\ \hline
    \end{tabular}
  \caption{Watershed level MCMC diagnostics are displayed. The effective final sample size of after thinning is computed as Effective $N_{final}$. The percentage of Geweke test Z-scores per watershed that are $<= |1.96|$ is displayed in the column labeled Non-Sig. Geweke Z-Scores.}    
     \label{mcmcdiagnosticstable}
\end{table}

%%%%%%%%%%%%%%%%%%%%%%%%%%%%%%%%%%%%%%%%%%%%%%%%%%%%%%%%%%%%%%%%%%%%%
% FIGURES
%%%%%%%%%%%%%%%%%%%%%%%%%%%%%%%%%%%%%%%%%%%%%%%%%%%%%%%%%%%%%%%%%%%%%
\clearpage

\newpage
\clearpage

\clearpage

\clearpage

\newpage
\clearpage

\newpage
\clearpage

\newpage
\clearpage

\newpage
\clearpage

\newpage
\clearpage

\newpage
\clearpage

\newpage
\clearpage

\newpage
\clearpage

\newpage
\clearpage

\newpage
\clearpage

\newpage
\clearpage

\newpage
\clearpage

\newpage
\clearpage

\newpage
\clearpage

\newpage
\clearpage

\newpage
\clearpage

\newpage
\begin{figure}[h]
\centering
	\includegraphics[width=\linewidth]{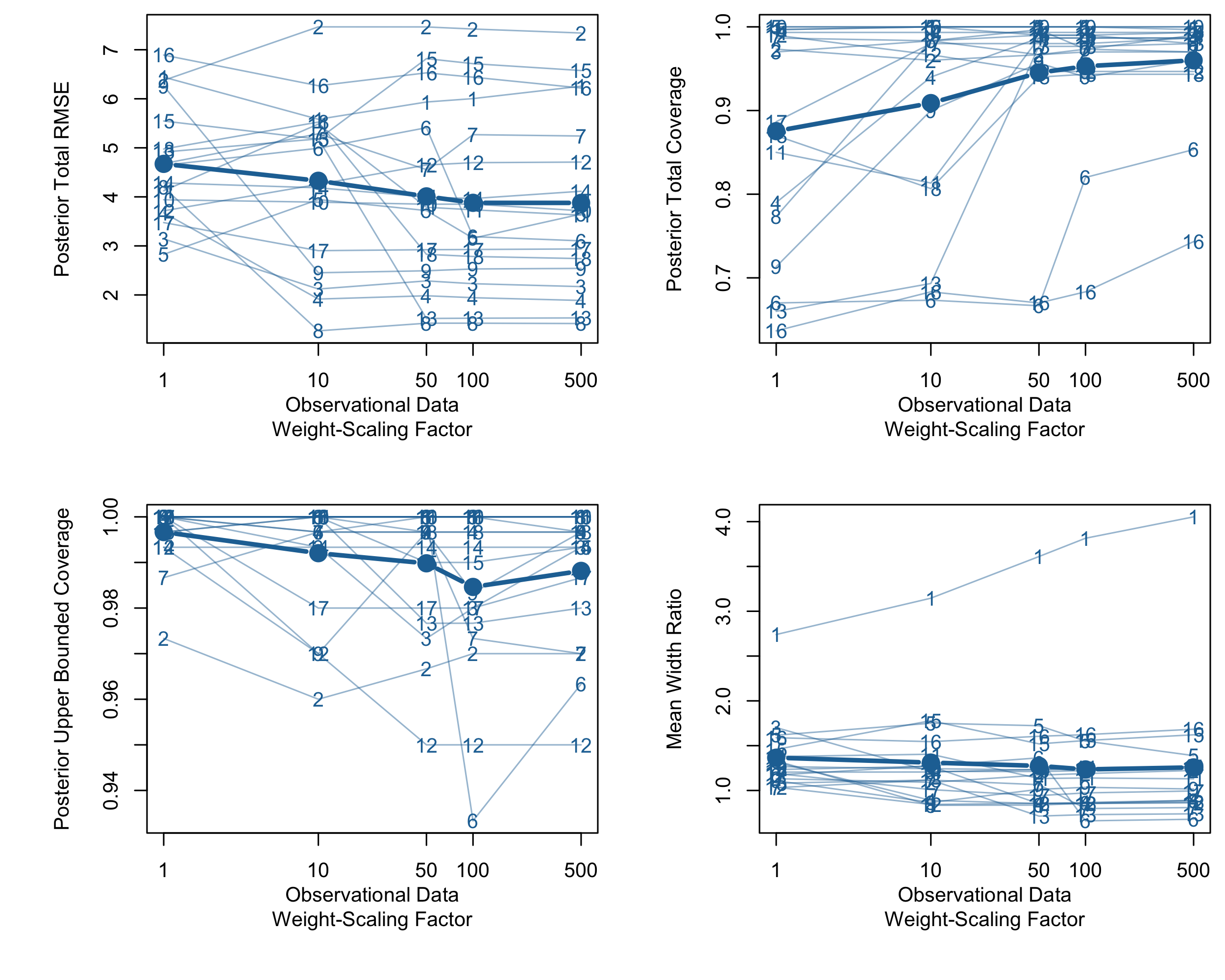}
	\caption{Posterior total RMSE ($RMSE_{p}$), posterior coverage ($cov_{p}$), upper posterior coverage ($cov^{u}_{p}$), and the ratio of average posterior distribution width over average ensemble width ($\frac{W_{p}}{W_{e}}$) are plotted as a function of $\tau_{k}$, which is set at values of 1, 10, 50, 100, and 500 (see Appendix section \ref{validation} for definitions on these metrics). For each subplot, opaque numbered blue lines represent individual watersheds, and the thicker blue line is the mean of those lines. Numbers map to watersheds from Table \ref{watersheds}. The quantities $cov_{p}$, $cov^{u}_{p}$, and $\frac{W_{p}}{W_{e}}$ are computed via a $99\%$ credible interval. $RMSE_{p}$ decreases until $\tau_{k}=100$. $cov_{p}$ increases but appears asymptote at $\tau_{k}=100$. $cov^{u}_{p}$ is generally insensitive to $\tau_{k}$ but does decrease slightly with larger values of $\tau_{k}$. With the exception of one watershed (Lower Mississippi Region, coded as 1 here), $\frac{W_{p}}{W_{e}}$ is insensitive to $\tau_{k}$.} 
	\label{ObservationalWeightScalingFactorExperimentResults}
\end{figure}
\clearpage

\newpage
\begin{figure}[h]
\centering
    \includegraphics[width=\linewidth]{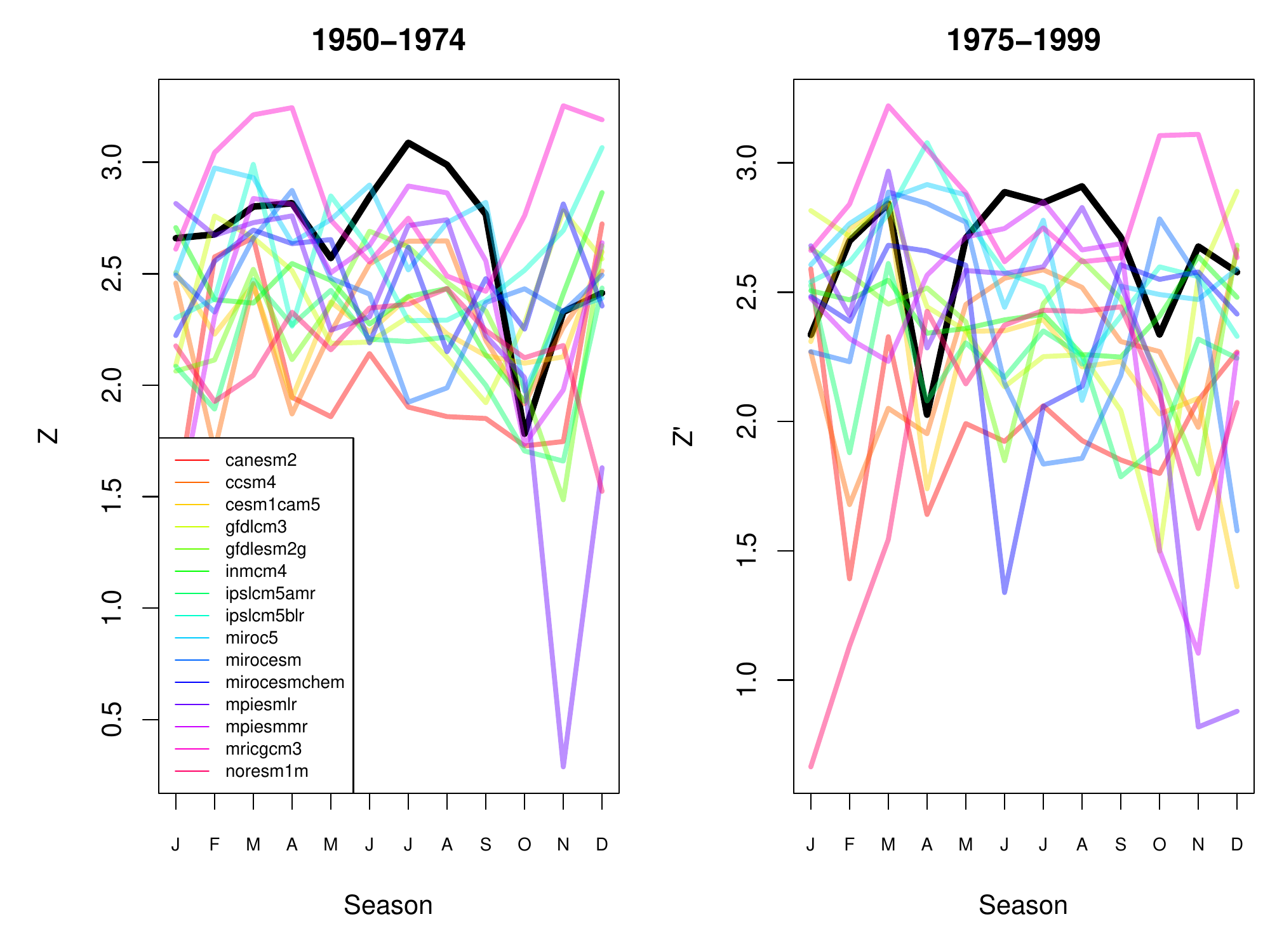}   
    \caption{(Log-transformed $q=1$ return levels are shown for observations ($Z_{k, m,q=1}$ and $Z'_{k,m,q'=1}$) and ESMs ($Z_{j,m,q=1}$ and $Z'_{j,m,q'=1}$) for the 1950-1974 and 1975-1999 climate regimes. The black line shows the observations and the colored lines the ESMs.}
    \label{q1plots}
\end{figure}
\clearpage

\newpage
\begin{figure}[h]
\centering
    \includegraphics[width=\linewidth]{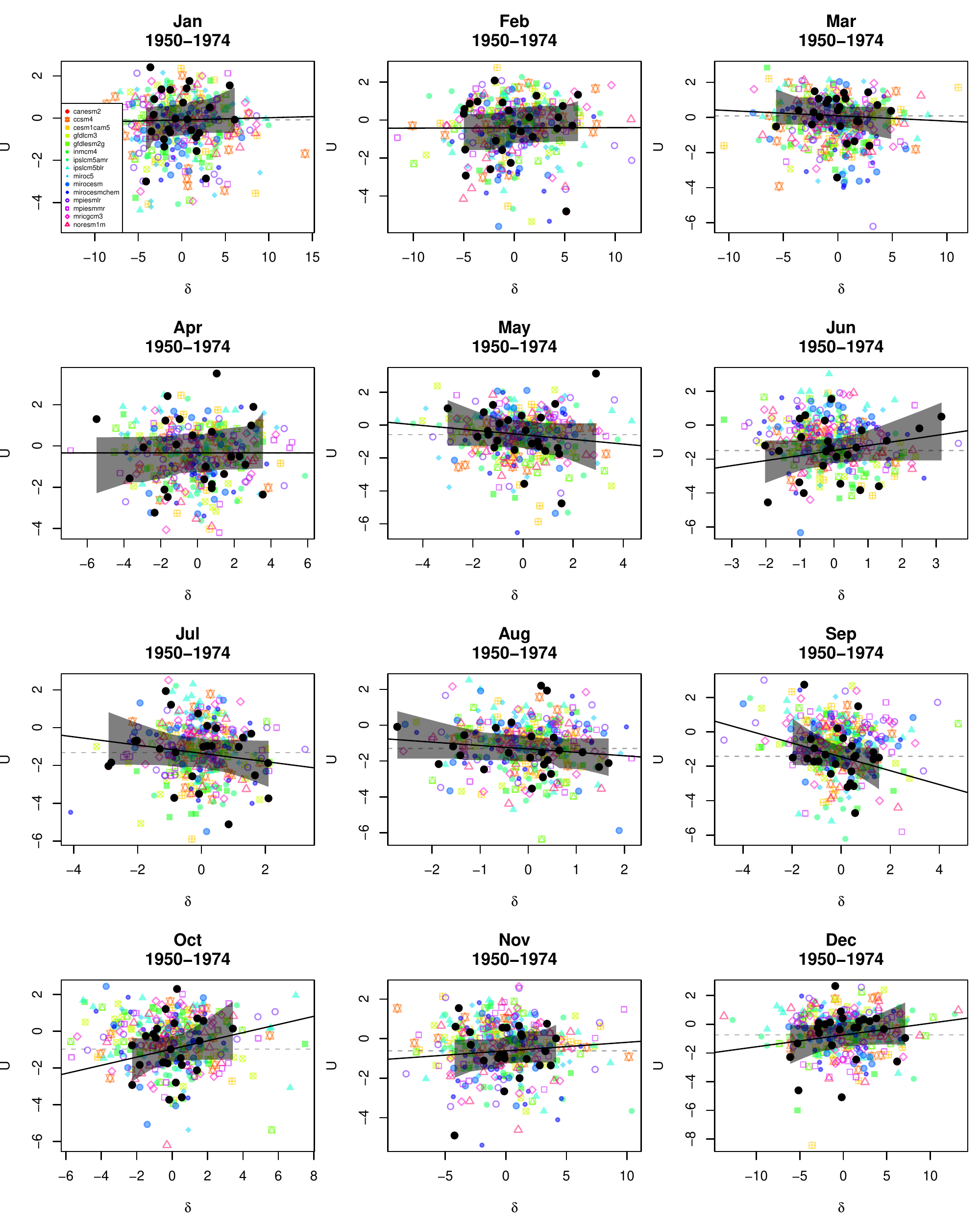}   
    \caption{For each month, $U_{j,m,q}$ are scatterplotted with corresponding values of $\delta_{j,m,q}$ for all datasets $j$ (and observations indexed by $k$). Black points are observations. Black lines and opaque bounds are ordinary least squares lines and $95\%$ prediction interval bounds, representing the observed relationship between same day temperature and precipitation return levels. Colored points represent ESMs, with each color and point type representing one ESM. A horizontal dashed line is shown at the mean of $U_{k,m,q}$ for context.}
    \label{qprimeplots19501974}
\end{figure}
\clearpage

\newpage
\begin{figure}[h]
\centering
        \includegraphics[width=\linewidth]{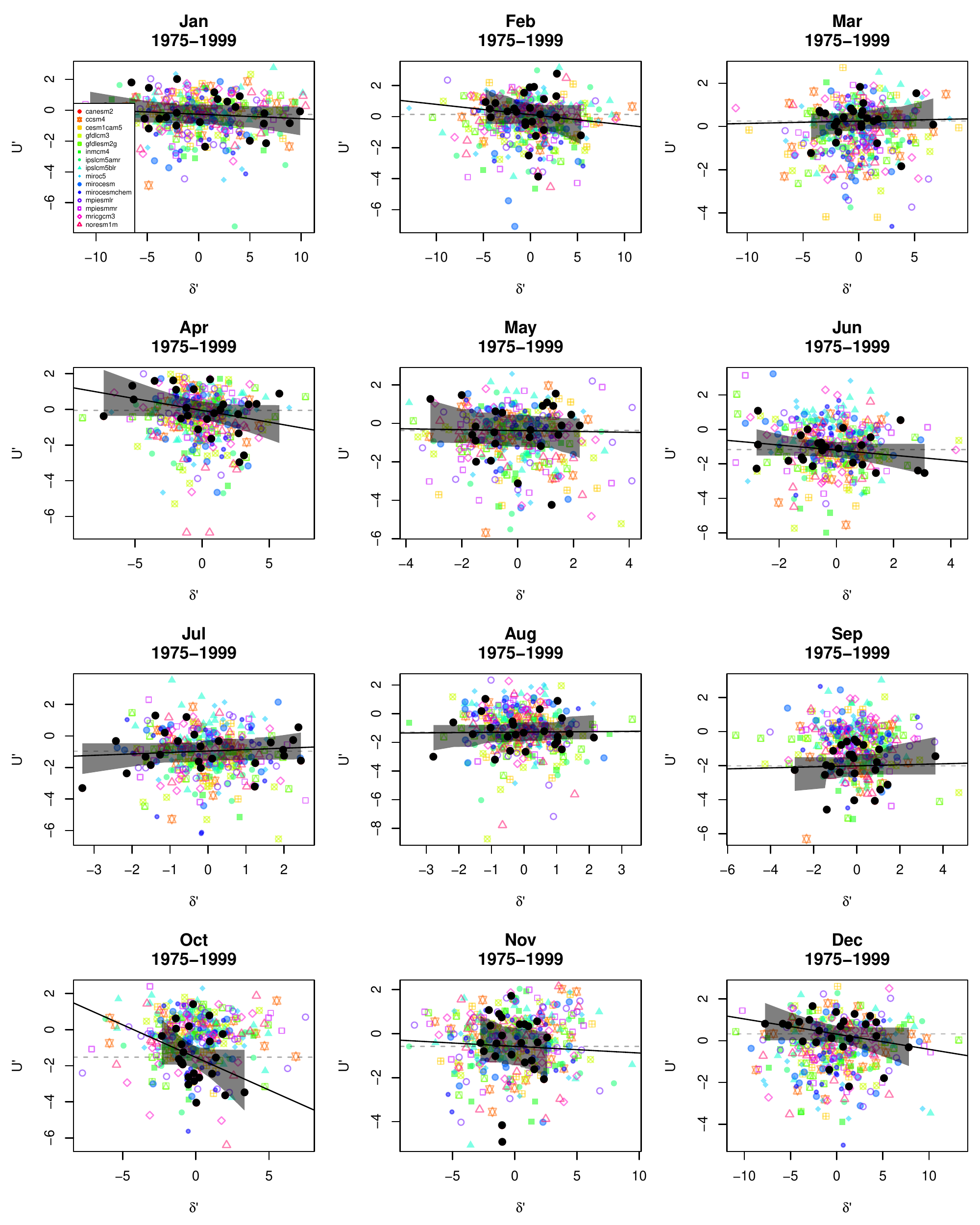} 
    \caption{The same as Figure \ref{qprimeplots19501974} is shown but for 1975-1999, i.e., $U'_{j,m,q'}$ and $\delta_{j,m,q'}$.}
    \label{qprimeplots19751999}
\end{figure}
\clearpage

\newpage
\begin{figure}[h]
\centering
        \includegraphics[width=0.80\linewidth]{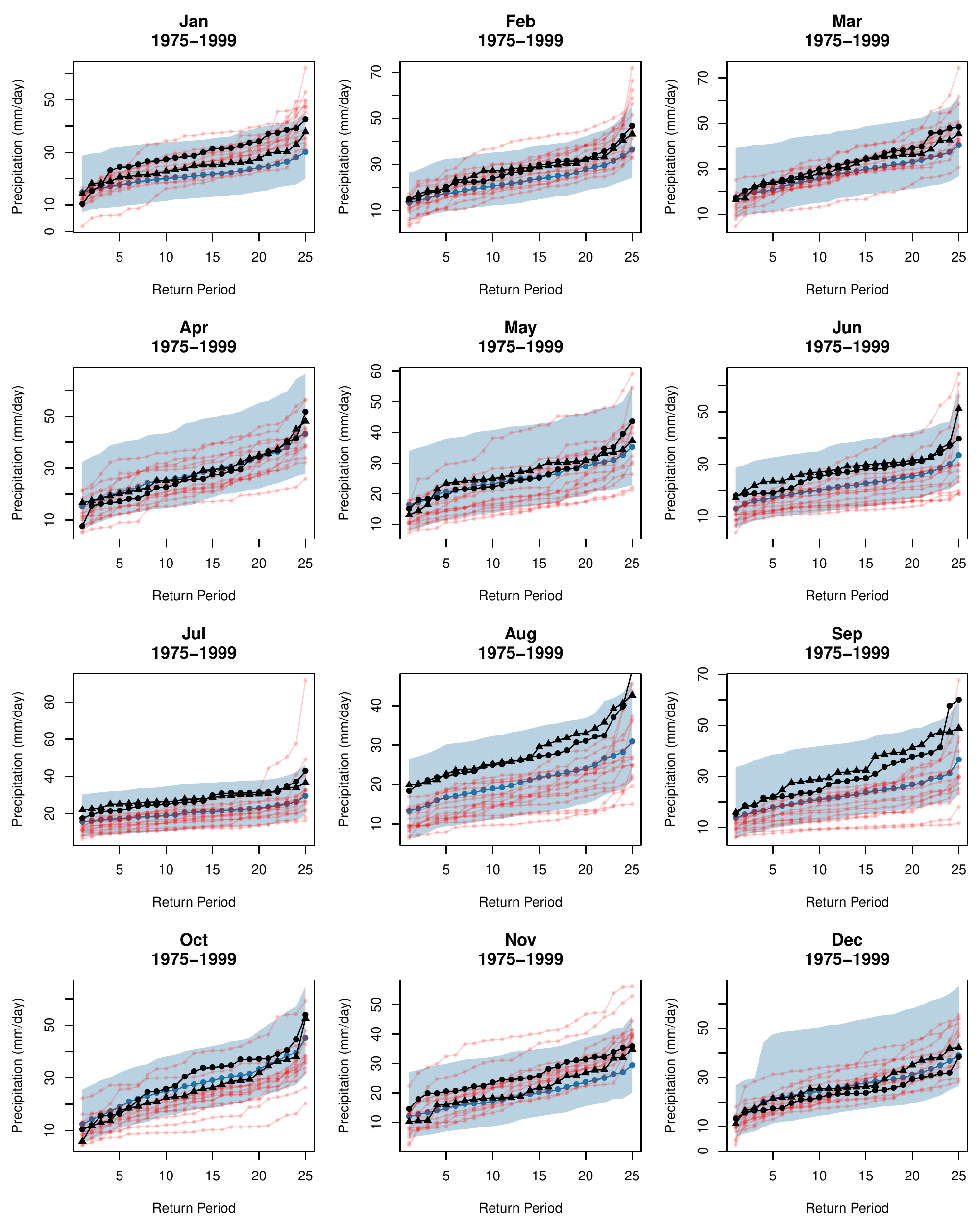} 
    \caption{Validation regime posterior distributions for $P'_{p, m, q', n \in [1,2,...N_{final}]}$ are shown for each month in the South Atlantic-Gulf watershed. Black dots are return levels of held out 1975-1999 observations, and black triangles are the same but from 1950-1974. Larger blue dots represent the posterior mean for each order statistic $q'$ (i.e., return period). Blue opaque bounds represent a $99\%$ credible interval for each $q'$ and $m$. Red points show original ESM return values.}
    \label{validationposteriors}
\end{figure}
\clearpage

\newpage
\begin{figure}[h]
\centering
        \includegraphics[width=\linewidth]{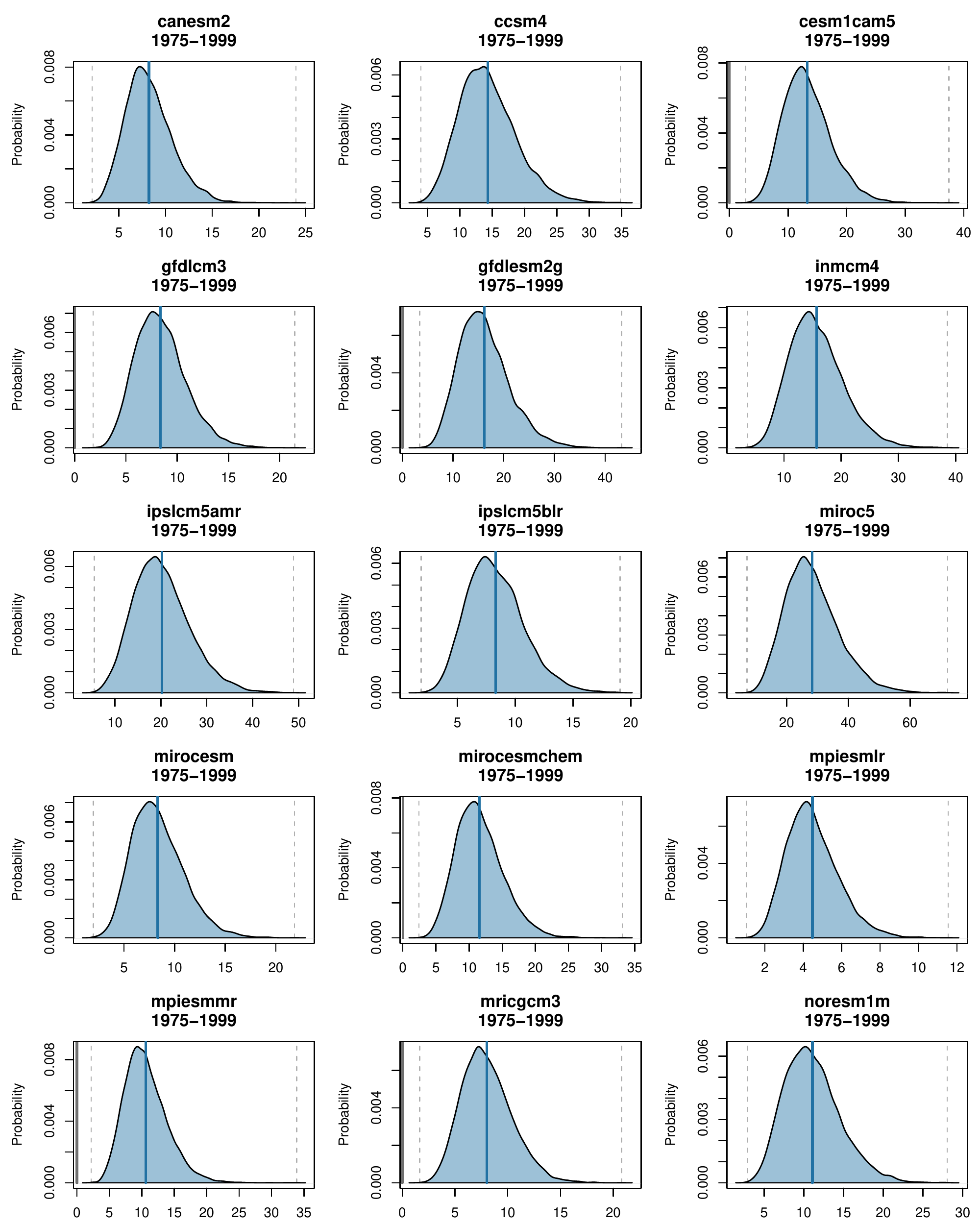} 
	\caption{Posteriors of $\sigma_{j}$ are shown for the validation scheme model run in the South Atlantic-Gulf watershed. Vertical blue lines indicate posterior means and dashed vertical gray lines show $99\%$ credible interval bounds.}
	\label{sigmaj}
\end{figure}
\clearpage

\newpage
\begin{figure}[h]
\centering
        \includegraphics[width=\linewidth]{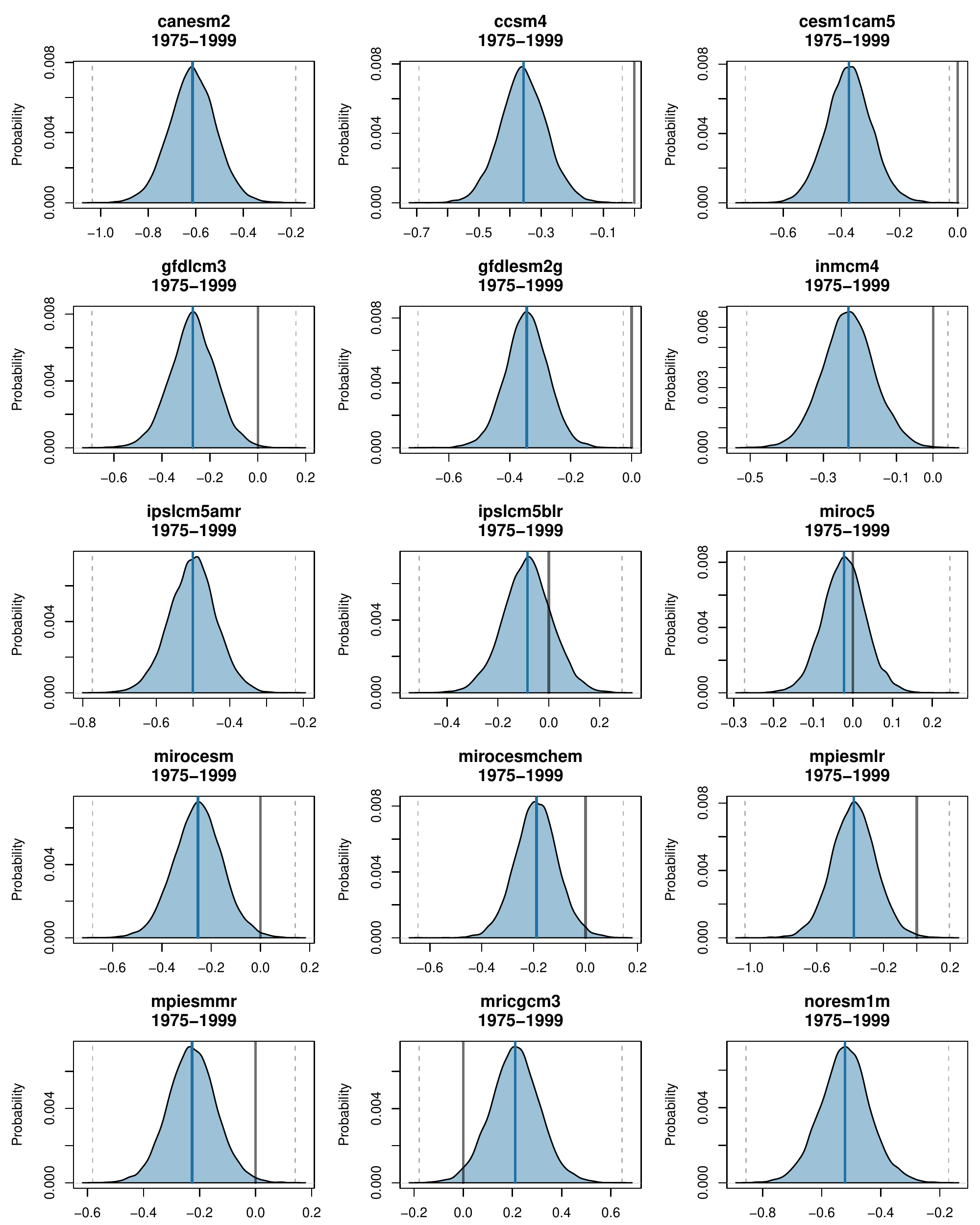} 
	\caption{Posteriors of $CBIAS_{j}$ are shown for the validation scheme model run in the South Atlantic-Gulf watershed. Vertical blue lines indicate posterior means and dashed vertical gray lines show $99\%$ credible interval bounds.}
	\label{cbiasj}
\end{figure}
\clearpage

\newpage
\begin{figure}[h]
\centering
        \includegraphics[width=0.80\linewidth]{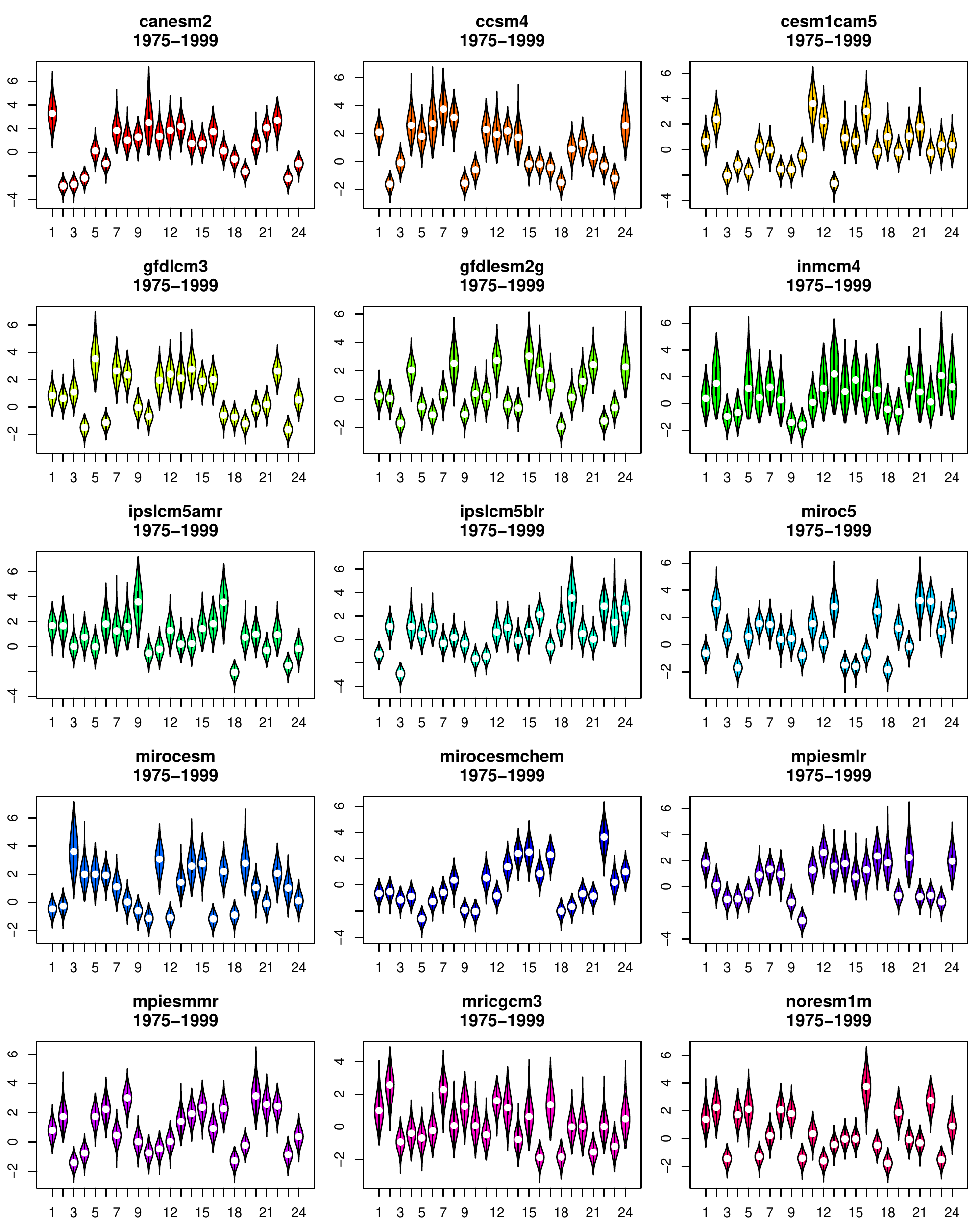} 
    \caption{Posteriors of $\epsilon_{j,q}$ (log scale) are depicted with violin plots for the validation scheme model run in the South Atlantic-Gulf watershed. Horizontal axes range from 1 to 24, which map to $q, q' \in [2, ... Q=Q'=25]$. The log scale is used to temper the visual effect of occasional large values. Values of $\epsilon_{j,q}$ can occasionally be large but typically smaller than $\epsilon_{k, q}$.}
    \label{epsilonjq}
\end{figure}
\clearpage
 
\newpage
\begin{figure}[h]
\centering
        \includegraphics[width=\linewidth]{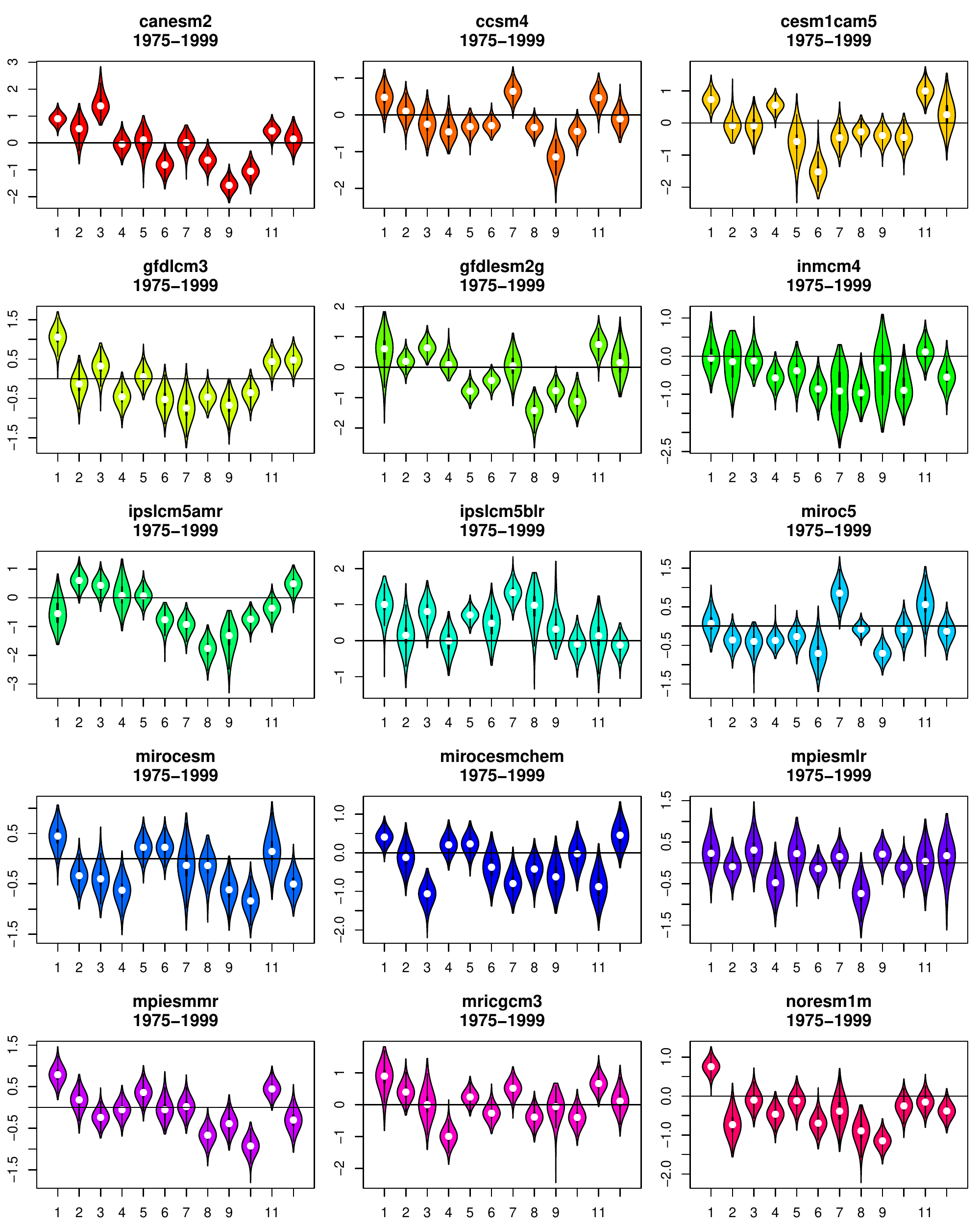} 
    \caption{Posteriors of $\alpha_{j,m}$ are shown via violin plots for the validation scheme model run in the South Atlantic-Gulf watershed. Horizontal axes range from 1 to 12, reflecting all 12 months.}
    \label{alphajm}
\end{figure}
\clearpage

\newpage
\begin{figure}[h]
\centering
        \includegraphics[width=\linewidth]{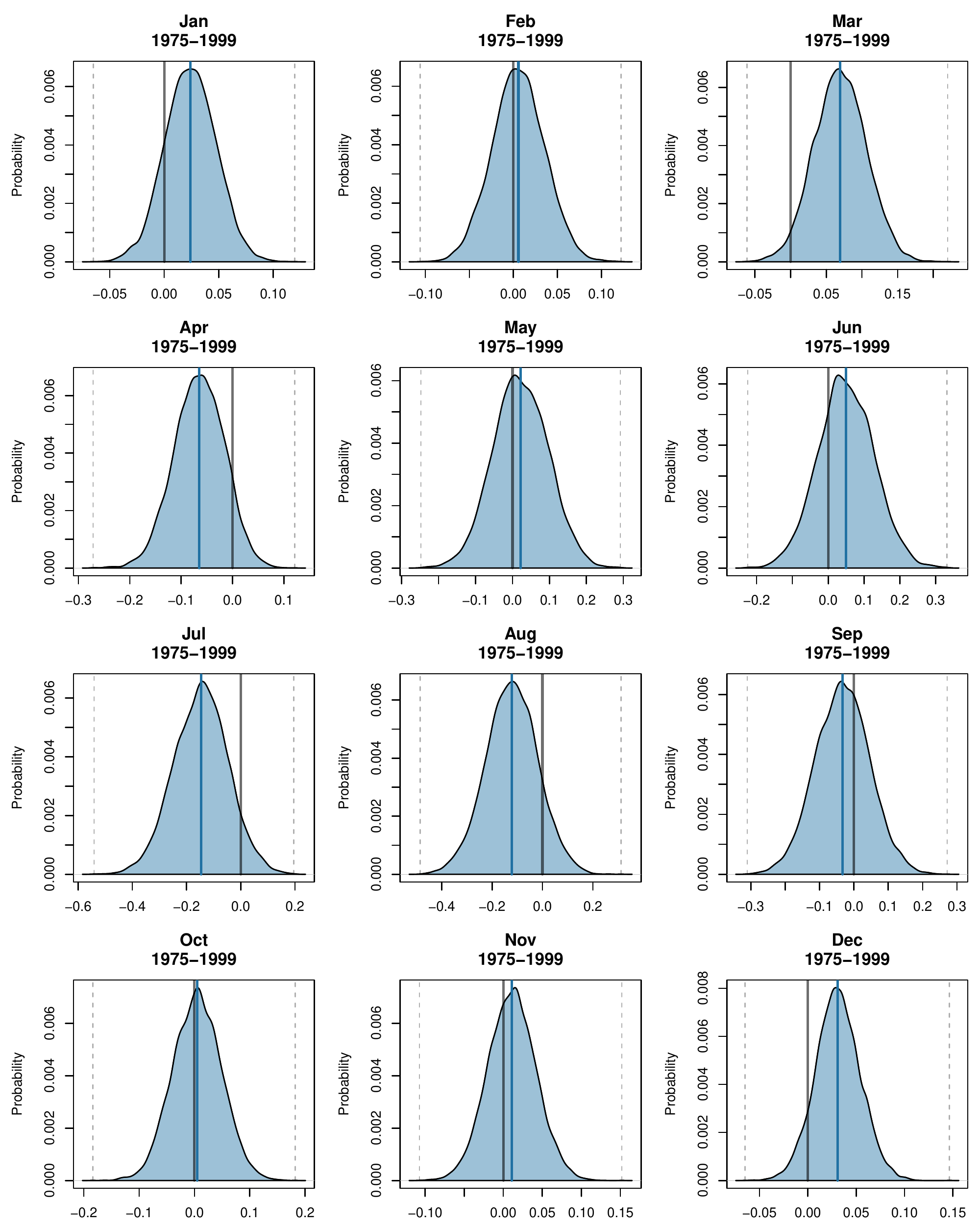} 
    \caption{Posteriors of $\phi'_{m}$ are shown for each month for the validation scheme model run  in the South Atlantic-Gulf watershed. Vertical blue lines represent the posterior mean, vertical solid gray lines are all at 0, and vertical dashed lines represent $99\%$ credible interval bounds.}
    \label{phiprime}
\end{figure}
\clearpage

\end{document}